\documentclass[12pt]{article}

\font\grande=cmr9.5 scaled \magstep4
\font\medio=cmr9.5 scaled \magstep2
\outer\def\beginsection#1\par{\medbreak\bigskip
      \message{#1}\leftline{\bf#1}\nobreak\medskip
\vskip-\parskip
      \noindent}
\usepackage{graphicx} 
\begin{document}
\bibliographystyle {unsrt}

\titlepage

\begin{flushright}
CERN-PH-TH/2011-017
\end{flushright}

\vspace{10mm}
\begin{center}
{\grande Gradient expansion, curvature perturbations}\\
\vskip 0.5cm
{\grande and magnetized plasmas}\\
\vspace{1.5cm}
 Massimo Giovannini$^{a,b}$\footnote{Electronic address: massimo.giovannini@cern.ch} and Zahra Rezaei$^{a,c}$\footnote{Electronic address: zahra.rezaei@cern.ch}\\
\vspace{1cm}
{{\sl $^{a}$Department of Physics, 
Theory Division, CERN, 1211 Geneva 23, Switzerland }}\\
\vspace{0.5cm}
{{\sl $^b$ INFN, Section of Milan-Bicocca, 20126 Milan, Italy}}\\
\vspace{0.5cm}
{{\sl $^c$ Isfahan University of Technology Department of Physics, 84154 Isfahan, Iran }}
\vspace*{0.5cm}
\end{center}

\vskip 0.5cm
\centerline{\medio  Abstract}
The properties of magnetized plasmas are always investigated under the hypothesis that the relativistic inhomogeneities stemming from the fluid sources and from the geometry itself are sufficiently small to allow for a perturbative description prior to photon decoupling. The latter assumption is hereby relaxed and pre-decoupling plasmas are described within a suitable expansion where the inhomogeneities are treated to a given order in the spatial gradients. It is argued that the (general relativistic) gradient expansion shares the same features of the drift approximation, customarily employed in the description of cold plasmas, so that the two schemes are physically complementary in the large-scale limit and for the low-frequency branch of the spectrum of plasma modes. The two-fluid description, as well as the magnetohydrodynamical reduction, are derived and studied in the presence of the spatial gradients of the geometry. Various solutions of the coupled system of evolution equations in the anti-Newtonian regime and in the quasi-isotropic approximation are presented. The relation of this analysis to the so-called separate Universe paradigm is outlined.  The evolution of the magnetized curvature perturbations in the nonlinear regime is addressed for the magnetized adiabatic mode in the plasma frame.
\noindent

\vspace{5mm}

\vfill
\newpage
\renewcommand{\theequation}{1.\arabic{equation}}
\setcounter{equation}{0}
\section{Motivations}
\label{sec1}
The analyses of the large-scale galaxy distribution \cite{LSS1,LSS2}, of the high-redshift type 
Ia supernovae \cite{SNN1,SNN2}  and of the 
Cosmic Microwave Background (CMB) observables \cite{WMAP7a,WMAP7b} seem to converge, these days, on a concordance model sometimes called $\Lambda$CDM scenario where $\Lambda$ stands for the 
dark energy component and CDM accounts for the dark matter component.  The $\Lambda$CDM 
scenario is just the compromise between the number of ascertainable parameters and the 
quality of the observational data.  The quest for a concordance lore is also able to shed some light on the presence of large-scale magnetic fields in nearly all gravitationally bound systems we observe.  Since we do see magnetic fields 
today over large distance scales, it seems natural to scrutinize their impact on the CMB observables. 
This is the motivation of a program aimed at bringing the unconventional study of magnetized CMB anisotropies 
to the same standard of the more conventional adiabatic\footnote{The wording ``adiabatic paradigm" refers 
here to the situation where the unique source of inhomogeneity prior to photon decoupling is localized in the 
standard adiabatic mode of curvature perturbations. This is the situation contemplated by the minimal 
version of the $\Lambda$CDM scenario.} paradigm (see \cite{mg1,mg2,mg3,mg4} and references therein). 
While different approaches to the problem are certainly available \cite{vt1,vt2,vt3,vt4,vt5} (see \cite{mg1a} for a more complete list of earlier references), the 
path followed in \cite{mg1,mg2} led to the calculation of the temperature and polarization anisotropies induced by the magnetized (adiabatic and entropic) initial conditions. 
The parameters of the magnetized background have been estimated (for the first time) in \cite{mg3,mg4} by using  the 
TT and TE correlations\footnote{Following the standard shorthand 
terminology the TT correlations denote the temperature 
autocorrelations while the TE correlations denote the cross-correlation between the temperature and the E-mode polarization.} measured by the WMAP collaboration. 
The obtained results\footnote{In \cite{mg1,mg2,mg3,mg4} the magnetic power 
spectrum and the magnetic spectral index $n_{\mathrm{B}}$ is defined with the conventions 
employed for the scalar modes of the geometry where the scale-invariant limit corresponds to $1$.}
 show that large-scale (comoving) magnetic fields larger than $3.5$ nG are excluded 
to 95 \% C.L. and for magnetic spectral indices $n_{\mathrm{B}} = 1.6_{-0.1}^{0.8}$.  These determinations 
have been conducted in the context of the minimal $m\Lambda$CDM where $m$ stands for magnetized. 
The addition of a fluctuating dark energy background pins down systematically larger values of the 
magnetic field  parameters (see \cite{mg4} for further details). 

The results obtained so far assumed the simplest setup for the inclusion 
of large-scale magnetic fields in the pre-decoupling plasma and it is therefore mandatory to scrutinize if the 
main assumptions of the analysis are consistently posited. For instance in \cite{mg1,mg2,mg3,mg4} 
(as well as in nearly all other approaches) it was assumed that magnetic fields do not contribute 
to the electron-photon scattering. In \cite{mg5} this assumption has been relaxed by explicitly including 
 the magnetic field contribution in the electron-photon scattering matrix.

In Refs. \cite{mg1,mg2,mg3,mg4}, as well as in other perturbative approaches 
to the physics of magnetized CMB anisotropies, a common hypothesis is that the 
intensity of the magnetized background is sufficiently small to describe magnetic fields 
within the standard Einstein-Boltzmann hierarchy where the curvature perturbations as well as 
the density contrasts all remain in the perturbative regime.  
In similar terms, perturbation theory is extremely well justified for the treatment 
of baryon acoustic oscillations because of the absolute smallness of the relative temperature 
fluctuations. As far as fully inhomogeneous magnetic fields are concerned, the perturbative assumption is a direct consequence of the closure bound applied to the magnetic field intensity.
The natural question, in this context, seems to be: can we 
go beyond the standard perturbative expansion and scrutinize its properties in a broader and firmer 
scheme? Can we understand which are the possible nonlinear effects in a gravitating and magnetized 
plasma to a given order in the spatial gradients? One of the standard tools to improve (and partially resum)
the perturbative description of the relativistic fluctuations of the geometry is the gradient expansion where 
the guiding criterion is not the absolute smallness 
of the given perturbation relative to its background value but rather the number 
of gradients carried by each term and defining the order of the expansion.  The extension 
of the general relativistic gradient expansion to the case of magnetized plasmas immediately suggests 
a physical connection with analog approximation schemes adopted in the discussion 
of weakly coupled plasmas in flat space-time.

The modest aim of the present analysis will then be to combine 
the general relativistic gradient expansion with the drift approximation (customarily applied in plasma physics) 
and to derive a system of equations where both expansions can be implemented in a unified manner.  
We remind that a known tool for computing the properties of cold plasma involves an expansion in the gradients 
of the magnetic and of the electric fields. The details of the scheme
depend upon the problem at hand and this richness is reflected in slightly different terminologies 
such as the drift approximation \cite{LL} (mainly adopted in kinetic theory) and the guiding center approximation \cite{ALF,BOY} (often used in the physics of cold plasmas and fully ionized gases\footnote{The expansion in spatial 
gradients is also relevant in nonlinear 
magnetohydrodynamics (MHD in what follows) in  flat space-time \cite{biskamp}.}). 

Similarly, the general relativistic gradient expansion has been discussed in several related contexts
since the papers of Lifshitz, Khalatnikov and Belinskii (see, e.g. \cite{lif1,lif2,BK1,KL}) but never in combination 
with either the drift approximation or the guiding centre approximation.  
More recently various applications of the gradient expansion to inflationary models have been studied
by Tomita \cite{tom1,tom2} as well as by Deruelle and collaborators \cite{dr1,dr2,dr3}.
The neglect of the spatial curvature and of the spatial gradients to zeroth order 
implies that the obtainable solutions are, in a sense, opposite 
to the ones customarily discussed in the Newtonian regime and this 
is the reason why they are sometimes named anti-Newtonian. The latter solutions 
are also relevant for the so-called separate Universe picture stipulating that any portion of the Universe larger than 
the Hubble radius $r_{\mathrm{H}}$ but smaller than the physical wavelength 
on the perturbation  will look like a separate unperturbed Universe.  
The gradient expansion has been also applied to the problem of the evolution 
of non-Gaussianities (see, e.g. \cite{tan1,tan2} and references therein) and to the problem 
of dark energy \cite{mg1g} with the purpose of demonstrating that the spatial gradients cannot be responsible 
of present-day accelaration (see e.g. \cite{wald} and references therein).  
An instructive approach to nonlinear power spectra in Einstein gravity has been 
developed by Noh, Hwang and collaborators in a series of interesting papers \cite{HN1,HN2,HN3}.
Finally the gradient expansion can be also employed in the investigation 
of the so-called sudden (or quiescent) singularities which arise in the context
of dark energy models with baotropic index smaller than the one of a cosmological constant (i.e. 
the supernegative equations of state) \cite{mg2g}. 

Even if some of the considerations developed in this paper will be rather general, it is useful to 
bear in mind the essentials of the pre-decoupling plasma which contains different 
components interacting both gravitationally and electromagnetically.  
To fix the notations, it is convenient  to separate the total action of the system, i.e. $S_{\mathrm{tot}}$ in three distinct parts
\begin{equation}
S_{\mathrm{tot}} = S_{\mathrm{grav}} + S_{\mathrm{em}} + S_{\mathrm{plasma}},
\label{ac1}
\end{equation}
where $S_{\mathrm{grav}}$ and $S_{\mathrm{em}}$ denote, respectively, 
the gravitational and the electromagnetic contributions
\begin{equation}
S_{\mathrm{grav}} + S_{\mathrm{em}} = \int \, d^{4} x \, \sqrt{ - g} \biggl[ -\frac{R}{2 \ell_{\mathrm{P}}^2} - \frac{1}{16\pi} 
F_{\alpha\beta}\, F^{\alpha\beta} - j_{\nu} A^{\nu} \biggr];
\label{ac2}
\end{equation}
note that $g= \mathrm{det} (g_{\mu\nu})$ and $F_{\alpha\beta}$ is the Maxwell field strength;
 $S_{\mathrm{plasma}}$ contains all the different components of the plasma which can be written, in 
the context of the vanilla $\Lambda$CDM paradigm, as 
\begin{equation}
S_{\mathrm{plasma}}= S_{\mathrm{e\,i}} + S_{\gamma} + S_{\nu} + S_{\mathrm{cdm}} + S_{\Lambda};
\label{ac3}
\end{equation}
$S_{\mathrm{e\,i}}$ denotes the contribution of electrons and ions; $S_{\gamma}$ and  $S_{\nu}$  
are, respectively, the contributions of the photons and of the (massless) neutrinos; $S_{\mathrm{cdm}}$ and 
$S_{\Lambda}$ account for the cold dark matter and for the dark energy.  

Before photon decoupling the condition of validity of the general relativistic 
gradient expansion and of the guiding center approximation are both verified and it is 
therefore extremely interesting to derive and study the evolution equations 
describing the plasma without assuming that the geometry is, a priori, 
conformally flat. Various questions can be addressed in this 
scheme such as, for instance, the corrections induced by the 
gradients of the geometry on the two-fluid plasma description, on the 
MHD reduction as well as on all the plasma processes which are 
relevant for a correct description of magnetized CMB anisotropies. Needless 
to say that  the present results are also applicable 
in the case of other magnetized systems in curved space-time not necessarily 
connected to CMB physics. 

The present paper is organized as follows. In section \ref{sec2} the decomposition of the inhomogeneous 
geometry will be introduced with special attention to the themes which are mostly relevant 
for the present discussion.  In section \ref{sec3} the evolution equations of  weakly coupled 
plasmas will be derived in the case where the metric is not assumed conformally flat and, 
in particular, without resorting to the standard separation between background 
geometry and relativistic fluctuations. In section \ref{sec4} we will scrutinize 
the way gradients must be counted in a magnetized plasma when the background 
geometry is fully inhomogeneous. 
Section \ref{sec5} discusses the anti-Newtonian solutions in the drift approximation. 
In section \ref{sec6} the quasi-isotropic MHD solutions are presented.  Section \ref{sec7} 
contains the derivation of the nonlinear magnetized 
adiabatic mode whose linearized counterpart is one of the elements 
of the simplest version of the $m\Lambda$CDM scenario.  Finally section \ref{sec8} 
contains our concluding remarks. In the appendix (divided in two parts, i.e. appendix \ref{APPA} and \ref{APPB}) 
the complementary results have been collected with the purpose 
of keeping the paper self-contained. 

\renewcommand{\theequation}{2.\arabic{equation}}
\setcounter{equation}{0}
\section{Fully inhomogeneous geometry}
\label{sec2}
The standard treatments of magnetized CMB anisotropies as well as 
the discussion of magnetized plasmas in the early Universe customarily assume that 
the geometry is separated into a homogeneous background supplemented
by its relativistic fluctuations. To go beyond the latter description,
 the fully inhomogeneous geometry shall then be described in terms of $N$, $N^{i}$ and $\gamma_{ij}$ 
denoting, respectively, the lapse function, the shift vector and the three-dimensional 
metric tensor, i.e.\footnote{The Greek indices will run over the 
four space-time dimensions while the Latin indices will denote the spatial 
indices.}  
\begin{eqnarray}
&& g_{00} = N^2 - N_{k} N^{k},\qquad g_{ij} = - \gamma_{ij},\qquad g_{0i} = - N_{i},
\nonumber\\
&& g^{00} = \frac{1}{N^2},\qquad g^{ij} = \frac{N^{i} \, N^{j}}{N^2}- \gamma^{ij},\qquad 
g^{0i} = - \frac{N^{i}}{N^2}.
\label{ADM1}
\end{eqnarray}
The decomposition of Eq. (\ref{ADM1}) is well known and it is sometimes 
referred to as the ADM decomposition from Arnowitt, Deser and Misner 
who applied it to the Hamiltonian formulation of the general relativistic dynamics (see, e.g., \cite{ADM1,ADM2}). 
In the ADM variables the extrinsic curvature 
$K_{ij}$ and the spatial components of the Ricci tensor $r_{ij}$ become:
\begin{eqnarray}
K_{ij} &=& \frac{1}{2 N} \biggl[- \partial_{\tau}\gamma_{ij} + ^{(3)}\nabla_{i}N_{j} + ^{(3)}\nabla_{j} N_{i} 
\biggr],
\label{ADM1a}\\
r_{ij} &=& \partial_{m} \, ^{(3)}\Gamma^{m}_{ij} -\partial_{j} ^{(3)}\Gamma_{i m}^{m} + ^{(3)}\Gamma_{i j}^{m} 
\,^{(3)}\Gamma_{m n}^{n} - ^{(3)}\Gamma_{j n}^{m} \,^{(3)}\Gamma_{i m}^{n},
\label{ADM1b}
\end{eqnarray}
where  $^{(3)}\nabla_{i}$ is the covariant derivative defined 
with respect to the metric $\gamma_{ij}$, $\partial_{\tau}$ denotes a derivation with respect to the time coordinate 
$\tau$ and $^{(3)}\Gamma_{i j}^{m}$ are the Christoffel symbols computed from $\gamma_{ij}$. Note that 
$\Gamma_{ij}^{m} = ^{(3)}\Gamma_{i j}^{m}$ but only in the case $N_{i}=0$ (see appendix \ref{APPA} for details). It is now useful to introduce the basic logic of the gradient 
expansion by looking at the whole system of Einstein equations with generic electromagnetic and fluid 
sources\footnote{As mentioned in section \ref{sec1} the species present in the plasma 
prior to photon decoupling can be interacting directly with the electromagnetic fields 
and this will demand a more detailed description of the sources as it will be discussed in section \ref{sec3}.}. In their contracted form the Einstein equations read
\begin{equation}
R_{\mu}^{\nu} = \ell_{\mathrm{P}}^2 \biggl[\biggl(T_{\mu}^{\nu} - \frac{T}{2} \delta_{\mu}^{\nu}\biggr) + {\mathcal T}_{\mu}^{\nu}\biggr], \qquad T= g^{\mu\nu} T_{\mu\nu} = T_{\mu}^{\mu},
\label{EE1}
\end{equation}
where $R_{\mu}^{\nu}$ is the four-dimensional Ricci tensor while $T_{\mu}^{\nu}$ and  ${\mathcal T}_{\mu}^{\nu}$
denote, respectively, the fluid energy-momentum tensor (accounting for the the global contribution of all the species 
of Eq. (\ref{ac3})) and the electromagnetic energy-momentum tensor:
\begin{equation}
T_{\mu\nu} = (p + \rho) u_{\mu} u_{\nu} - p g_{\mu\nu}, \qquad {\mathcal T}_{\mu}^{\nu} = \frac{1}{4\pi} \biggl( - F_{\mu \alpha} F^{\nu\alpha} + \frac{\delta_{\mu}^{\nu}}{4} F_{\alpha\beta} F^{\alpha\beta} \biggr),
\label{EE2}
\end{equation}
where  $g^{\mu\nu} u_{\mu} u_{\nu} =1$ (see also Eq. (\ref{T2}) of appendix \ref{APPA} 
for an explicit expression of the latter condition in the ADM metric (\ref{ADM1})).
If not otherwise stated the covariant derivatives with spatial 
indices listed below will always refer to the three-dimensional metric. In other words, 
to simplify the notation, $^{(3)} \nabla_{i} \equiv \nabla_{i}$. Details on the inhomogeneous 
geometry of Eq. (\ref{ADM1}) can be found in appendix \ref{APPA} and will be 
quoted whenever needed.  Bearing in mind these conventions,  the various components 
of Eq. (\ref{EE1}) are given by:
\begin{eqnarray}
&& \partial_{\tau} K - N \mathrm{Tr} K^2 + \nabla^2 N = N \ell_{\mathrm{P}}^2 \biggl\{ \frac{3 p + \rho}{2} 
 + ( p + \rho) \, u^2 + {\mathcal T}_{0}^{0}\biggr\},
\label{EE3}\\
&& \nabla_{i} K - \nabla_{k} K^{k}_{i} = N \ell_{\mathrm{P}}^2 \biggl[ \frac{u_{i}}{N} ( p + \rho) \sqrt{1 + u^2} + {\mathcal T}_{i}^{0}
\biggr],
\label{EE4}\\
&& \partial_{\tau} K_{i}^{j} - N K K_{i}^{j} - N r_{i}^{j} + \nabla_{i} \nabla^{j} N = \ell_{\mathrm{P}}^2 N\biggl[ \frac{p - \rho}{2} \delta_{i}^{j} - ( p + \rho) u_{i} u^{j} +  {\mathcal T}_{i}^{j} \biggr],
\label{EE5}
\end{eqnarray}
where, for sake  of simplicity,  the shorthand notation $u^2 = \gamma^{ij} u_{i} u_{j}$ has been adopted. 
Having chosen $N_{i}=0$ in the general equations of appendix \ref{APPA}, the geometry appearing in Eqs. (\ref{EE3})--(\ref{EE5}) is described in terms of $7$ independent functions (i.e. $\gamma_{ij}$ and $N$).  
It is useful to keep the lapse function arbitrary for practical purposes such as, for instance, 
the matching with the perturbative treatment of the magnetized fluctuations in the conformally Newtonian gauge 
(see, e.g., appendix \ref{APPB}).  The momentum constraint (i.e. Eq. (\ref{EE4})) can be used, in the framework of the general relativistic gradient expansion, to obtain the velocity field in terms of the extrinsic curvature evaluated to the 
preceding order in the expansion. By combining the trace of Eq. (\ref{EE5}) with Eq. (\ref{EE3}), the standard form of the Hamiltonian constraint can be readily obtained
\begin{equation}
K^2 - \mathrm{Tr}K^2 + r = 2 \ell_{\mathrm{P}}^2 \biggl[ \rho + {\mathcal T}_{0}^{0} + 
(p + \rho) u^2 \biggr].
\label{EE6}
\end{equation}
The traceless part of Eq. (\ref{EE5}) can instead be written as:
\begin{equation}
\partial_{\tau} \overline{K}_{i}^{j} - N K \overline{K}_{i}^{j} - N \overline{r}_{i}^{j} + \nabla_{i} \nabla^{j} N - \frac{\nabla^2 N}{3} \delta_{i}^{j} = \ell_{\mathrm{P}}^2 N \biggl\{ - (p + \rho) \biggl[ u_{i} u^{j} - \frac{u^2}{3} \delta_{i}^{j}\biggr] + \overline{{\mathcal T}}_{i}^{j}\biggr\},
\label{EE7}
\end{equation}
where the barred quantities define the traceless part of the corresponding variable, i.e.
\begin{equation}
\overline{K}_{i}^{j} = K_{i}^{j} - \frac{K}{3} \delta_{i}^{j}, \qquad \overline{r}_{i}^{j} = r_{i}^{j} - \frac{r}{3} \delta_{i}^{j}, \qquad 
\overline{{\mathcal T}}_{i}^{j} = {\mathcal T}_{i}^{j} - \frac{{\mathcal T}}{3} \delta_{i}^{j}.
\label{EE8}
\end{equation}
Recalling Eq. (\ref{ADM6}) and using Eq. (\ref{EE8}) it follows that $\mathrm{Tr} K^2 = \overline{K}_{i}^{j} \, \overline{K}_{j}^{i} + K^2/3$. Bearing in mind the explicit form of Eqs. (\ref{EE3})--(\ref{EE5}),  to zeroth order in the spatial gradients, the peculiar velocities as well as the spatial curvature are neglected. From the momentum constraint, the zeroth-order results 
determine the peculiar velocity which can be used as an input for the following order in the expansion. By iteration the first-order correction to the geometry can be determined. While the zeroth order of
 the linearized approximation is by definition homogeneous, the zeroth-order of the gradient expansion can well be fully inhomogeneous but does not contain any spatial gradient. Conversely, the first-order depends upon the spatial gradients and upon the spatial curvature and so on and so forth.
 
If the gravitating system is a magnetized plasma (such as the one present prior to photon decoupling) 
the inhomogeneities of the electromagnetic fields will necessarily modify the trajectories 
of the charged species. In this case the nature of the physical system combines 
inextricably electromagnetic and gravitational gradients. The tenets of the (general relativistic) gradient expansion 
must be consistently combined with the expansion in spatial derivatives usually adopted in the analysis of weakly 
coupled plasmas in flat space-time \cite{ALF,BOY}  both in the two- and one-fluid approximations. There are different ways of introducing the guiding centre approximation and the simplest one is to think of a gradient expansion of the large-scale magnetic field, i.e. denoting  with $\vec{B}$ the (flat-space) 
magnetic field we can write that
\begin{equation}
B_{i}(\vec{x},\tau) \simeq B_{i}(\vec{x}_{0}, \tau)  + (x^{j} - x_{0}^{j}) \partial_{j} B_{i} +...
\label{G1}
\end{equation}
where the ellipses stand for the higher orders in the gradients leading, both, to curvature and drift corrections. A similar expansion can also be written in the case of the electric 
field with the caveat that, in a plasma, electric fields are anyway screened for typical length-scale 
larger than the Debye radius:
\begin{equation}
E_{i}(\vec{x},\tau) \simeq E_{i}(\vec{x}_{0}, \tau)  + (x^{j} - x_{0}^{j}) \partial_{j} E_{i} +...
\label{G2}
\end{equation}
In a flat-space plasma to zeroth-order in the expansion of Eqs. (\ref{G1}) and (\ref{G2}) only the time derivative of the magnetic fields are kept. To first-order the spatial derivatives of the magnetic field can then be used as 
an input to deduce  the electric fields. The first derivatives of the electric fields (obtained to first-order)
can be used to deduce the second spatial derivatives of the magnetic fields and so on.
If we ought to combine the general relativistic gradient expansion with the drift approximation,
the essential step is the  generalization of the two-fluid description and of the usual MHD reduction to the situation where the 
fully inhomogeneous geometry is parametrized as in Eq. (\ref{ADM1}). This analysis will lead automatically 
to the correct dynamical variables whose explicit form can be compared, for instance, with the corresponding 
variables deduced in the context of conformally flat geometries which are usually assumed in more conventional perturbative expansions. 
\newpage
\renewcommand{\theequation}{3.\arabic{equation}}
\setcounter{equation}{0}
\section{Charged fluids and gradient expansion}
\label{sec3}
 The electromagnetic interaction affects the evolution of the electron-photon-ion system while it affects only indirectly the evolution of the weakly interacting species. Since large-scale magnetic fields gravitate, the relativistic fluctuations of the geometry are modified by their presence via Eqs. (\ref{EE3}), (\ref{EE4}) and (\ref{EE5}). 
The purpose of the present section is to deduce the usual magnetohydrodynamical 
(MHD) reduction to lowest order in the gradient expansion, i.e. by allowing an underlying 
geometry with potentially large inhomogeneities over sufficiently large length-scales. The obtained 
equations generalize previous perturbative results  (see, in particular, \cite{mg3}) and will be cross-checked  
in the appropriate limits.
The Maxwell equations in a four-dimensional curved space-time can be written as 
\begin{equation}
\nabla_{\mu} F^{\mu\nu} = 4 \pi j^{\nu},\qquad \nabla_{\mu} \tilde{F}^{\mu\nu} =0,
\label{EL1}
\end{equation}
where $\nabla_{\mu}$ is the covariant derivative defined with respect 
to the four-dimensional metric $g_{\mu\nu}$ while $\tilde{F}^{\mu\nu}$ denotes the dual 
field strength. In terms of the ADM decomposition of Eq. (\ref{ADM1}) the field strengths and their duals are: 
\begin{eqnarray}
F^{0 i} &=& - \frac{{\mathcal E}^{i}}{N^2}, \qquad F^{ij} = - \gamma_{m k} \, \eta^{i j k} \,\frac{{\mathcal B}^{m}}{N},
\label{EL2}\\
\tilde{F}^{0 i} &=& - \frac{{\mathcal B}^{i}}{N^2},\qquad 
\tilde{F}^{ij} = \gamma_{m k} \,\eta^{i j k} \,\frac{{\mathcal E}^{m}}{N}.
\label{EL3}
\end{eqnarray}
The totally antisymmetric Levi-Civita tensor $\eta^{i j k}$ is defined as
\begin{equation}
\eta^{i j k} = \frac{\epsilon^{ i j k}}{\sqrt{\gamma}}, \qquad \eta_{i j k} = \sqrt{\gamma}\, 
\epsilon_{i j k},
\label{EL4a}
\end{equation}
where $\epsilon_{i j k}$ is the Levi-Civita symbol in flat space.
The Maxwell field strengths with doubly covariant indices are given in appendix \ref{APPA} (see Eq.  (\ref{T18}))
and simplify when the shift vector vanishes.
Defining with $e$ the electron charge, the total current appearing in Eq. (\ref{EL1}) is the sum of the currents of the electrons and of the ions\footnote{In Eq. (\ref{EL5a}) as well as in other equations the (lowercase) roman subscripts 
label the species (for instance electrons or ions) while the (lowercase) italic subscripts (or superscripts) must be understood 
as (spatial) tensor indices.}
\begin{equation}
j^{\nu} = j_{\mathrm{(e)}}^{\nu} +  j_{\mathrm{(i)}}^{\nu}, \qquad 
j^{\nu}_{(\mathrm{e})} = - e \tilde{n}_{\mathrm{e}} \, u_{\mathrm{e}}^{\nu}, \qquad j^{\nu}_{(\mathrm{i})} =  e
\tilde{n}_{\mathrm{i}}\,u_{\mathrm{i}}^{\nu},
\label{EL5a}
\end{equation} 
where $\tilde{n}_{\mathrm{e}}$ and $\tilde{n}_{\mathrm{i}}$ denote, respectively, the 
concentrations of electrons and ions.  The generic four-velocity $u^{\mu}$ satisfies 
$g^{\mu\nu} u_{\mu} u_{\nu} = 1$ which means that 
\begin{equation}
u^{0} = \frac{\cosh{y}}{N}, \qquad u^{i}   = \frac{v^{i}\,\cosh{y}}{N},\qquad \cosh{y} = \frac{1}{\sqrt{1 - \frac{v^2}{N^2}}},
\label{EL6a}
\end{equation}
where $v^2 = \gamma_{ij} v^{i} \, v^{j}$.  Note that, from now on, we shall deal, without loss of generality 
with the case $N_{i}= 0$.  For immediate convenience 
the notations of Eq. (\ref{EL6a}) can also be recast in a slightly different form by defining 
$\hat{v}^{i} = v^{i}/v$:
\begin{equation}
u^{i} = \hat{v}^{i} \, \sinh{y},\qquad u_{i} u^{j} = \hat{v}_{i} \hat{v}^{j} \sinh^2{y},\qquad u^{i} u_{0} = \hat{v}^{i} \, N\,\cosh{y} \sinh{y},
\label{EL7a}
\end{equation}
where $\hat{v}^{i} \hat{v}^{j} \gamma_{ij} =1$. Bearing in mind Eqs. (\ref{EL5a})--(\ref{EL7a}), 
Eq. (\ref{EL1}) demands the validity of the following system of equations:
\begin{eqnarray}
&& \partial_{i} \biggl[  \frac{\sqrt{\gamma}}{N} {\mathcal E}^{i} \biggr] = 4 \pi e [n_{\mathrm{i}} - n_{\mathrm{e}}], \qquad \partial_{i} \biggl[  \frac{\sqrt{\gamma}}{N} {\mathcal B}^{i} \biggr]=0,
\label{EL4}\\
&& \partial_{\tau} \biggl[ \frac{\sqrt{\gamma}}{N} {\mathcal B}^{i}\biggr] + 
\partial_{j}\biggl[ \eta^{j k i}\, \sqrt{\gamma}\, \gamma_{k \ell} {\mathcal E}^{\ell} \biggr] =0,
\label{EL5}\\
&& \partial_{j}\biggl[ \sqrt{\gamma}\, \eta^{j k i} \, {\mathcal B}^{m} \gamma_{k m} \biggr] = 4 \pi e \biggl[n_{\mathrm{i}} \, v_{(\mathrm{i})}^{i} - n_{\mathrm{e}} \, v_{(\mathrm{e})}^{i}\biggr] + \partial_{\tau} \biggl[ \frac{\sqrt{\gamma}}{N} {\mathcal E}^{i}\biggr],
\label{EL6}
\end{eqnarray}
where the electron and ion concentrations have been rescaled as:
\begin{equation}
n_{(\mathrm{i})} = \tilde{n}_{\mathrm{i}}\, \sqrt{\gamma}\, \cosh{y_{\mathrm{i}}}, \qquad 
n_{(\mathrm{e})} = \tilde{n}_{\mathrm{e}}\, \sqrt{\gamma} \cosh{y_{\mathrm{e}}}.
\label{EL8}
\end{equation}
From the covariant conservation of the electron and ion currents  (i.e. 
$\nabla_{\mu} j^{\mu}_{(\mathrm{e})} =0$ and $\nabla_{\mu} j^{\mu}_{(\mathrm{i})} =0$),
the evolution equations for the rescaled electron and ion concentrations 
\begin{equation}
\partial_{\tau} n_{\mathrm{e}} 
+ \partial_{k} [ n_{\mathrm{e}} v^{k}_{\mathrm{e}}] =0,\qquad 
\partial_{\tau} n_{\mathrm{i}}  
+ \partial_{k} [ n_{\mathrm{i}} v^{k}_{\mathrm{i}}] =0,
\label{EL16}
\end{equation}
imply that, to lowest order in the gradient expansion, $n_{\mathrm{e}}$ and $n_{\mathrm{i}}$ 
are constant.  Equations (\ref{EL4}), (\ref{EL5}) and (\ref{EL6}) assume a simpler form by 
an appropriate rescaling of the electric and magnetic fields:
\begin{eqnarray}
&& \vec{\partial} \cdot \vec{E} = 4 \pi e [n_{\mathrm{i}} - n_{\mathrm{e}}], 
\qquad \vec{\partial} \cdot \vec{B}=0,
\label{EL10}\\
&& \partial_{\tau} \vec{B} + \vec{\partial} \times \vec{E}  =0,
\label{EL11}\\
&& \vec{\partial} \times \vec{B} = 4 \pi e \biggl[n_{\mathrm{i}} \, \vec{v}_{\mathrm{i}} - n_{\mathrm{e}} \, \vec{v}_{\mathrm{e}}\biggr] + \partial_{\tau}\vec{E}, 
\label{EL12}
\end{eqnarray}
where the rescaled electric and magnetic fields are given by:
\begin{equation}
E^{i} = \frac{\sqrt{\gamma}}{N} {\mathcal E}^{i},\qquad B^{i} = \frac{\sqrt{\gamma}}{N} {\mathcal B}^{i}.
\label{EL9}
\end{equation}
The shorthand notation employed in Eqs. (\ref{EL10}), (\ref{EL11}) and (\ref{EL12}) implies
for a generic vector $A^{i}$, 
\begin{equation} 
\vec{\partial}\cdot \vec{A} \equiv \partial_{i} A^{i},\qquad 
(\vec{\partial}\times \vec{A})^{i} = \partial_{j}\biggl[N \gamma^{i k} \, \gamma^{j n} \, \eta_{n m k} \, A^{m}\biggr].
\label{EL12a}
\end{equation}
The vectors appearing in Eqs. (\ref{EL10}), (\ref{EL11}) and (\ref{EL12}) 
become three-dimensional Cartesian vectors  in the limit when the background geometry is 
homogeneous, isotropic and conformally flat, i.e.
\begin{equation}
N \to a(\tau), \qquad \gamma_{ i j} = a^{2}(\tau) \delta_{i j}.
\label{EL14}
\end{equation}
Using Eq. (\ref{EL14}) into Eq. (\ref{EL8}) we have that, for $y_{\mathrm{e}} \ll 1$ and $y_{\mathrm{i}}\ll1 $, the 
electron and ion concentrations\footnote{Electrons and ions are non-relativistic across matter-radiation equality 
and around decoupling.  
Still, because of the masses of the electrons and ions, the conformal invariance of the whole 
system is broken (see, e.g. \cite{mg1,mg2}).}  become
$n_{\mathrm{e}} = a^3 \tilde{n}_{\mathrm{e}}$ and $n_{\mathrm{i}} = a^3 \tilde{n}_{i}$. 
Using Eq. (\ref{EL14}) into  Eqs.  (\ref{EL10})--(\ref{EL12})  the standard differential operators 
are recovered, i.e.  $\vec{\partial} \to \vec{\nabla}$, $ \vec{\partial} \times \vec{A} \to 
\vec{\nabla} \times \vec{A}$.  In general terms, the system is not conformally invariant as it can be argued by looking at the form of Eq. (\ref{EL12a}) 
and by appreciating that indices are raised and lowered in terms of $\gamma_{ij}$. Various 
discussions obtained in the limit defined by Eq. 
(\ref{EL14}) \cite{mg3,mg4} (see also \cite{mg1,mg2}) can be generalized to the fully inhomogeneous 
situation. Without dwelling on all possible generalizations we shall focus our attention only on
a consistent implementation of the two-fluid and one-fluid descriptions.

Neglecting, for a moment, the terms responsible for the momentum exchange between electrons and ions the covariant conservation of the energy momentum tensor of the charged species implies 
\begin{equation}
 \nabla_{\mu} T^{\mu\nu}_{(\mathrm{e})} = j^{(\mathrm{e})}_{\alpha} F^{\nu\alpha},
\qquad 
\nabla_{\mu} T^{\mu\nu}_{(\mathrm{i})} = j^{(\mathrm{i})}_{\alpha} F^{\nu\alpha},
\label{EL20}
\end{equation}
where 
\begin{equation}
T^{\mu\nu}_{(\mathrm{e})} =  \rho_{\mathrm{e}} u^{\mu}_{\mathrm{e}} u^{\nu}_{\mathrm{e}} ,\qquad  
T^{\mu\nu}_{(\mathrm{i})} = \rho_{\mathrm{i}} u^{\mu}_{\mathrm{i}} u^{\nu}_{\mathrm{i}}.
\label{EL22}
\end{equation}
Using Eq. (\ref{T9}) and choosing the free index of Eq. (\ref{EL20}) to be time-like, the explicit 
evolution of the energy density of the electrons can be obtained and it is:
\begin{eqnarray}
&& \partial_{\tau}[ \rho_{\mathrm{e}} \cosh^2{y_{\mathrm{e}}}] +  N^2 \partial_{k} \biggl[ 
\frac{\rho_{\mathrm{e}}}{N} \cosh{y_{\mathrm{e}}} \sinh{y_{\mathrm{e}}} \hat{v}_{\mathrm{e}}^{k}\biggr] 
\nonumber\\
&&- N K \rho_{\mathrm{e}} \cosh^2{y_{\mathrm{e}}}  + \biggl[3 \nabla_{k} N + N \Gamma_{k j}^{j}\biggr]\rho_{\mathrm{e}} 
\cosh{y_{\mathrm{e}}} \sinh{y_{\mathrm{e}}} \hat{v}_{\mathrm{e}}^{k} 
\nonumber\\
&& - N K_{k j} \biggl[ \rho_{\mathrm{e}} \hat{v}_{\mathrm{e}}^{k} \, \hat{v}_{\mathrm{e}}^{j} \sinh^2{y_{\mathrm{e}}}
+ \Pi_{\mathrm{e}}^{k j}\biggr]
 = - j^{(\mathrm{e})}_{k} {\mathcal E}^{k},
\label{EL24}
\end{eqnarray}
where the contribution of the anisotropic stress of the electrons $\Pi^{ij}_{\mathrm{e}}$ has been also included 
for completeness. The same expression holds also in the case of the ions by flipping the sign of the electric charge 
and by replacing $\mathrm{e} \to \mathrm{i}$ in the various 
subscripts (i.e. $\rho_{\mathrm{e}} \to \rho_{\mathrm{i}}$, $y_{\mathrm{e}} \to y_{\mathrm{i}}$, ... and so on and so forth).
By choosing the free index of Eq. (\ref{EL20}) to be space-like the resulting equation is
\begin{eqnarray}
&& \partial_{\tau}[ \rho_{\mathrm{e}} \cosh{y_{\mathrm{e}}} \sinh{y_{\mathrm{e}}} \hat{v}^{k}_{\mathrm{e}}] + 
\partial_{m}[ \rho_{\mathrm{e}} \sinh^2{y_{\mathrm{e}}} \, \hat{v}^{m}_{\mathrm{e}} \hat{v}^{k}_{\mathrm{e}} + \Pi_{\mathrm{e}}^{m k}]
\nonumber\\
&& - 2 N K_{m}^{k} \rho_{\mathrm{e}} \sinh{y_{\mathrm{e}}} \cosh{y_{\mathrm{e}}} \hat{v}^{m}_{\mathrm{e}} + 
\nabla^{k} N \rho_{\mathrm{e}} \cosh^2{y_{\mathrm{e}}} - N K \rho_{\mathrm{e}} \sinh{y_{\mathrm{e}}} \cosh{y_{\mathrm{e}}}
\hat{v}^{k}_{\mathrm{e}} 
\nonumber\\
&&+ [ \nabla_{m} N + N \Gamma_{n m}^{n}] [ \rho_{\mathrm{e}} \hat{v}^{m}_{\mathrm{e}} \hat{v}^{k}_{\mathrm{e}} \sinh^2{y_{\mathrm{e}}} + \Pi^{k m}_{\mathrm{e}}] + N \Gamma_{m n}^{k} [ \rho \hat{v}^{m} \hat{v}^{n} s^2(y) +  \Pi^{m n}] 
\nonumber\\
&&= j_{0}^{(\mathrm{e})} \frac{{\mathcal E}^{k}}{N} - \gamma_{m q} j_{n}^{(\mathrm{e})} \eta^{k n m}
{\mathcal B}^{q} + {\mathcal C}_{\mathrm{e}\, i} + {\mathcal C}_{\mathrm{e\, \gamma}},
\label{EL25}
\end{eqnarray}
where ${\mathcal C}_{\mathrm{e}\, i}$ and ${\mathcal C}_{\mathrm{e\, \gamma}}$ denote the 
collision terms of the electrons with ions and photons. Similarly the evolution 
equation for the ion velocity field can be obtained by replacing $\mathrm{e} \to \mathrm{i}$ 
in the relevant subscripts in full analogy with what has been already suggested, after Eq. (\ref{EL24}), 
for the evolution equations of the energy density. 
Equations (\ref{EL24}) and (\ref{EL25}) can be expanded in gradients and, to lowest order, the evolution of the electron and ion 
energy densities can be derived from Eq. (\ref{EL24}):
\begin{equation}
\partial_{\tau} \rho_{\mathrm{e}} - N K \rho_{\mathrm{e}} = - e n_{\mathrm{e}} \frac{\vec{v}_{\mathrm{e}} \cdot \vec{E}}{\gamma}, \qquad\partial_{\tau} \rho_{\mathrm{i}} - N K \rho_{\mathrm{i}} = e n_{\mathrm{i}} \frac{\vec{v}_{\mathrm{i}} \cdot \vec{E}}{\gamma},
\label{EL27}
\end{equation}
where, generically, the following notation will be employed throughout the paper:
\begin{equation}
\vec{F} \cdot \vec{G} = \gamma_{m n} F^{m} G^{n},\qquad (\vec{F} \times \vec{G})^{k} = \frac{\gamma_{i n} \gamma_{m \ell}}{N}
F^{n} G^{m} \eta^{i \,\ell\, k}.
\label{EL31}
\end{equation}
The reduction from the two-fluid  to the one-fluid description follows the standard steps of the flat-space time case (see, e.g. \cite{spitzer}) but with the difference that the terms stemming from the fully inhomogeneous 
nature of the underlying geometry will be consistently taken into account.  The idea of the one-fluid reduction 
is to pass from a description  holding for the two (or more) separate species to a one-fluid 
system where the dynamical variables are global, like the total current, the baryon 
energy density, the center-of-mass velocity of the electron-ion system and so on and so forth. 
The sum of the ion and of the electron energy densities is usually referred to as the baryon density and 
its evolution equation is obtained by summing, term by term, the two equations appearing in Eq. (\ref{EL27}):	
\begin{equation}
\partial_{\tau} \rho_{\mathrm{b}} - N K \rho_{\mathrm{b}} = \frac{\vec{J} \cdot \vec{E}}{\gamma}, 
\label{EL28}
\end{equation} 
where $\rho_{\mathrm{b}}$ and $\vec{J}$ are, respectively, the baryon density and the total current:
\begin{equation}
\rho_{\mathrm{b}} = \rho_{\mathrm{e}} + \rho_{\mathrm{i}}, \qquad \vec{J} = e (n_{\mathrm{i}} \vec{v}_{\mathrm{i}} - 
n_{\mathrm{e}} \vec{v}_{\mathrm{e}}).
\label{EL29}
\end{equation}
In Eq. (\ref{EL28}), the contribution of the electric field has been kept since it multiplies 
the total current which will turn out to be related with spatial gradients of the magnetic fields as well as with 
the spatial gradients of the geometry possibly multiplied by terms containing the magnetic field itself.  From Eq. (\ref{EL25}) the evolution equation for the electron velocity can be written as:
\begin{eqnarray}
\partial_{\tau} v_{\mathrm{e}}^{k} + N \partial^{k} N - {\mathcal G}^{k}_{j} v_{\mathrm{e}}^{j} &=& - \frac{e \tilde{n}_{\mathrm{e}} N^2}{ \rho_{\mathrm{e}} \sqrt{\gamma}} \biggl[ E^{k} + (\vec{v}_{\mathrm{e}} \times \vec{B})^{k} \biggr] 
\nonumber\\
&+& N \Gamma_{\mathrm{ei}} (v_{\mathrm{i}}^{k} - v_{\mathrm{e}}^{k}) + \frac{4}{3} 
\frac{\rho_{\gamma}}{\rho_{\mathrm{e}}} N \Gamma_{\mathrm{e} \gamma}(v_{\gamma}^{k} - v_{\mathrm{e}}^{k}),
\label{EL30}
\end{eqnarray}
where 
\begin{equation}
 {\mathcal G}^{k}_{j}= \biggl[\frac{\partial_{\tau} N}{N} \delta_{j}^{k} + 2 N K_{j}^{k}\biggr].
\label{EL30A}
\end{equation} 
In Eq. (\ref{EL30}) the collision terms have been included and the shorthand notation of Eq. (\ref{EL31}) has been used
for the vector product. Similarly, the evolution equation for the ion velocity is given by
\begin{eqnarray}
\partial_{\tau} v_{\mathrm{i}}^{k} + N \partial^{k} N - {\mathcal G}^{k}_{j} v_{\mathrm{i}}^{j} &=& \frac{e \tilde{n}_{\mathrm{i}} N^2}{ \rho_{\mathrm{i}} \sqrt{\gamma}} \biggl[ E^{k} + (\vec{v}_{\mathrm{i}} \times \vec{B})^{k} \biggr] 
\nonumber\\
&+& N \Gamma_{\mathrm{ie}} (v_{\mathrm{e}}^{k} - v_{\mathrm{i}}^{k}) + \frac{4}{3} \frac{\rho_{\gamma}}{\rho_{\mathrm{i}}} N \Gamma_{\mathrm{i} \gamma}(v_{\gamma}^{k} - v_{\mathrm{i}}^{k}).
\label{EL32}
\end{eqnarray}
By summing up Eq. (\ref{EL30}) (multiplied by the electron mass) and Eq. (\ref{EL32}) (multiplied by the ion mass) 
 the evolution equation for center of mass velocity of the electron-ion system 
\begin{equation}
v_{\mathrm{b}} = \frac{m_{\mathrm{e}} v_{\mathrm{e}}^{k} + m_{\mathrm{i}} v_{\mathrm{i}}^{k}}{m_{\mathrm{e}} + m_{\mathrm{i}}}
\label{EL33a}
\end{equation}
can be obtained and the evolution equations for the baryon-lepton-photon system are
\begin{eqnarray}
\partial_{\tau} \rho_{\gamma} &=& \frac{4}{3} K N \rho_{\gamma} - \frac{4}{3} N \partial_{k}\biggl( \frac{\rho_{\gamma}}{N}\,v_{\gamma}^{k}\biggr),
\label{EL33}\\
\partial_{\tau} v_{\mathrm{b}}^{k} &=& {\mathcal G}_{j}^{k} v_{\mathrm{b}}^{j}  - N \partial^{k} N  + \frac{(\vec{J} \times \vec{B})^{k} N^2}{ \gamma\, \rho_{\mathrm{b}} ( 1 + m_{\mathrm{e}}/m_{\mathrm{i}})} + 
\frac{4}{3} \frac{\rho_{\gamma}}{\rho_{\mathrm{b}}} N \Gamma_{\gamma\mathrm{e}} (v_{\gamma}^{k} - v_{\mathrm{b}}^{k}), 
\label{EL34}\\
\partial_{\tau} v_{\gamma}^{k} &=& \biggl[ {\mathcal G}_{j}^{k} - \frac{N K}{3} \delta_{j}^{k}\biggr] v_{\gamma}^{j} - \frac{N^2}{4 \rho_{\gamma}} \partial_{m} \biggl(\rho_{\gamma} \gamma^{m k} \biggr) - N \partial^{k} N +
N \Gamma_{\gamma \mathrm{e}} ( v_{\mathrm{b}}^{k} - v_{\gamma}^{k}),
\label{EL35}
\end{eqnarray}
where $v_{\gamma}^{k}$ and $\rho_{\gamma}$ denote, respectively, the photon velocity and the photon 
energy density. The possibility of describing the baryon-lepton-photon fluid 
as a unique physical entity is a direct consequence of the fact that 
the electron-ion collision rate is much larger than the electron-photon (or ion-photon) 
rate.  While the electron-photon rate increases with the temperature the 
Coulomb rate decreases. The meeting point of the two rates occurs 
close to the MeV. Equations (\ref{EL33})--(\ref{EL35}) can then be used 
below the meeting point of the two rates.

Equation (\ref{EL32}) (multiplied by $e\, n_{\mathrm{i}}$) can be subtracted from Eq. (\ref{EL30}) (multiplied by $e n_{\mathrm{e}}$) leading to the generalized Ohm equation, i.e. the evolution equation of the total current:
\begin{eqnarray}
&& \frac{\partial J^{k}}{\partial \tau} - {\mathcal G}_{j}^{k} J^{j}  = - e \biggl[ v^{k}_{\mathrm{i}} \partial_{j} ( n_{\mathrm{i}} v_{\mathrm{i}}^{j}) - v_{\mathrm{e}}^{k} \partial_{j} (n_{\mathrm{e}} v_{\mathrm{e}})\biggr] 
- e (n_{\mathrm{i}} - n_{\mathrm{e}}) N \partial^{k} N
\nonumber\\
&&+ \frac{\omega_{\mathrm{pe}}^2 + \omega_{\mathrm{pi}}^2}{4\pi} \vec{E} 
- e n_{\mathrm{i}} \frac{N \partial_{j}[ p_{\mathrm{i}} \gamma^{k j}]}{ \rho_{\mathrm{i}}} + 
e n_{\mathrm{e}} \frac{N \partial_{j}[ p_{\mathrm{e}} \gamma^{k j}] }{ \rho_{\mathrm{e}}}
\nonumber\\
&& + e n_{\mathrm{e}} \, N\,\Gamma_{\mathrm{ei}} \biggl(1 + \frac{m_{\mathrm{e}}}{
m_{\mathrm{i}}}\biggr)\biggr[ \frac{(n_{\mathrm{i}} - n_{\mathrm{e}}) (m_{\mathrm{e}} + m_{\mathrm{i}})}{m_{\mathrm{i}} n_{\mathrm{e}} + n_{\mathrm{i}} m_{\mathrm{e}}} v_{\mathrm{b}}^{k} - \frac{(m_{\mathrm{i}} + m_{\mathrm{e}})}{e (n_{\mathrm{i}}
 m_{\mathrm{e}} + n_{\mathrm{e}} m_{\mathrm{i}})} J^{k} \biggr]
\nonumber\\
&& + \frac{e^2 n_{\mathrm{e}} n_{\mathrm{i}}(m_{\mathrm{e}} + m_{\mathrm{i}}) N^2}{ 
 \sqrt{\gamma} m_{\mathrm{e}}(n_{\mathrm{i}} m_{\mathrm{e}} + n_{\mathrm{e}} m_{\mathrm{i}})}\biggl(1 + \frac{m_{\mathrm{e}}}{m_{\mathrm{i}}}\biggr) (\vec{v}_{\mathrm{b}} \times \vec{B})^{k} 
\nonumber\\
&& + \frac{e N^2}{(n_{\mathrm{i}} m_{\mathrm{e}} + n_{\mathrm{e}} m_{\mathrm{i}}) \sqrt{\gamma}}
\biggl(n_{\mathrm{i}} \frac{m_{\mathrm{e}}}{m_{\mathrm{i}}} - n_{\mathrm{e}}
\frac{m_{\mathrm{i}}}{m_{\mathrm{e}}}\biggr) (\vec{J}\times \vec{B})^{k}
\nonumber\\
&& + \frac{4}{3} e \rho_{\gamma} \biggl\{ \biggl( \frac{\Gamma_{\mathrm{i}\gamma}}{m_{\mathrm{i}}} - 
\frac{\Gamma_{\mathrm{e} \gamma}}{m_{\mathrm{e}}}\biggr) v_{\gamma}^{k} 
+ \biggl[ \frac{\Gamma_{\mathrm{e}\gamma} n_{\mathrm{i}} (m_{\mathrm{e}} + m_{\mathrm{i}})}{m_{\mathrm{e}} ( 
m_{\mathrm{i}} n_{\mathrm{e}} + m_{\mathrm{e}} n_{\mathrm{i}})} - 
\frac{\Gamma_{\mathrm{i}\gamma} n_{\mathrm{e}} (m_{\mathrm{e}} + m_{\mathrm{i}})}{m_{\mathrm{i}} ( 
m_{\mathrm{i}} n_{\mathrm{e}} + m_{\mathrm{e}} n_{\mathrm{i}})} \biggr] v_{\mathrm{b}}^{k}
\nonumber\\
&& - \biggl[ \frac{\Gamma_{\mathrm{e}\gamma}}{e (m_{\mathrm{i}} n_{\mathrm{e}} + 
n_{\mathrm{i}} m_{\mathrm{e}})} \biggl(\frac{m_{\mathrm{i}}}{m_{\mathrm{e}}}\biggr) + 
\frac{\Gamma_{\mathrm{i}\gamma}}{e (m_{\mathrm{i}} n_{\mathrm{e}} + 
n_{\mathrm{i}} m_{\mathrm{e}})} \biggl(\frac{m_{\mathrm{e}}}{m_{\mathrm{i}}}\biggr)\biggr] J^{k}\biggr\},
\label{OHM1}
\end{eqnarray}
where Eq. (\ref{EL16}) have been used and where the plasma  frequencies for electrons and ions are defined, respectively,  as 
\begin{equation}
\omega_{\mathrm{p\, e}}^2 = \frac{4 \pi e^2 n_{\mathrm{e}} N^2}{m_{\mathrm{e}} \sqrt{\gamma}}, \qquad 
\omega_{\mathrm{p\, i}}^2 = \frac{4 \pi e^2 n_{\mathrm{i}} N^2}{m_{\mathrm{i}} \sqrt{\gamma}}.
\label{OHM2}
\end{equation}
Since the plasma is globally neutral,  $n_{\mathrm{e}} = n_{\mathrm{i}} = n_{0}$ (with\footnote{Recall that $\eta_{\mathrm{b}}$ denotes the 
ratio between the baryonic concentration and the photon concentration and it is given by 
$\eta_{\mathrm{b}}=6.219 \times 10^{-10} (h_{0}^2\Omega_{\mathrm{b}0}/0.02773) (2.725\, \mathrm{K}/T_{\gamma 0})^{3}$ where $h_{0}$ is the indetermination on the present value of the Hubble rate and $\Omega_{\mathrm{b}0}$ parametrizes the 
present value of the critical fraction of baryons; $T_{\gamma 0}$ is the CMB temperature.} $n_{0} = \eta_{\mathrm{b}} n_{\gamma}$). Thus Eq. (\ref{OHM2}) greatly simplifies also because of the 
smallness of the ratio between the electron and ion mass:
\begin{eqnarray}
&& \frac{\partial J^{k}}{\partial \tau} + \biggl[ N \Gamma_{\mathrm{ie}} \delta_{j}^{k}+ \frac{4 \rho_{\gamma} \Gamma_{\mathrm{e}\gamma}}{3 n_{0} m_{\mathrm{e}}} \delta_{j}^{k}- {\mathcal G}_{j}^{k}\biggr] J^{j}  = - e \biggl[ v^{k}_{\mathrm{i}} \partial_{j} ( n_{\mathrm{i}} v_{\mathrm{i}}^{j}) - v_{\mathrm{e}}^{k} \partial_{j} (n_{\mathrm{e}} v_{\mathrm{e}})\biggr] 
\nonumber\\
&&+ \frac{\omega_{\mathrm{p\, e}}^2}{4\pi}\biggl[ E^{k} + \biggl(\vec{v}_{\mathrm{b}}\times \vec{B}\biggr)^{k} + \frac{N}{e n_{0}} \partial_{j}\biggl(p_{\mathrm{e}} \gamma^{k j}\biggr) - \frac{(\vec{J} \times \vec{B})^{k}}{e n_{0}}\biggr]. 
\label{OHM3}
\end{eqnarray}
In Eq. (\ref{OHM3}) the terms containing the spatial gradients 
have been kept to illustrate the analogies and the differences 
with the customary MHD discussions in flat space-time. The third and fourth terms appearing in the 
square bracket of the second line of Eq. (\ref{OHM3}) are, respectively, the thermoelectric term 
(containing the gradient of the pressure) and the Hall term (containing the vector product of the current and of the 
magnetic field).  The thermoelectric term contains pressure gradients and it is therefore 
of higher order in the gradient expansion. The terms containing the velocity field multiplied by its own gradient 
are even smaller than thermoelectric term. The Hall term
is of the same order of the thermoelectric term.
Equation (\ref{OHM3}) can also be schematically written as
\begin{equation}
\partial_{\tau} J^{k} + \biggl(  N \Gamma \delta_{i}^{k} - {\mathcal G}_{i}^{k} \biggr) J^{i} = S^{k},
\label{Red1}
\end{equation}
where $\Gamma = \Gamma_{\mathrm{i\, e}} + [ 4 \rho_{\gamma}/(3 n_{0} m_{\mathrm{e}})] \Gamma_{\mathrm{e}\gamma})$ denotes the sum of the electron-ion and electron photon rates.  Equations (\ref{EL34}) 
and (\ref{EL35}) can be combined by noticing  that, for sufficiently early times prior to decoupling,
the baryon and photon velocity coincide. The resulting equation generalizes the standard evolution 
equation for the baryon-photon velocity which plays a role in the semi-analytic treatment 
of the magnetized CMB anisotropies at small angular scales \cite{mg3}. Another possible 
generalization concerns the propagation of electromagnetic waves 
in the plasma which is relevant for the Faraday rotation of the polarization plane of the CMB (see, e.g. \cite{FAR}).
In conclusion all the evolution equations customarily employed for the description of weakly coupled plasmas
in the linearized approximation can be generalized to the case of a fully inhomogeneous geometry expressed 
in ADM variables.

The evolution equations derived in this section reproduce, in the conformally flat limit, the 
standard perturbative results of Ref. \cite{mg3}. As an example consider 
Eqs. (\ref{EL33}), (\ref{EL34}) and (\ref{EL35}) describe the evolution 
of the photon-baryon system to lowest order in the gradient expansion. All the quantities 
appearing in the latter equations depend both on $\vec{x}$ and $\tau$ and generalize 
the perturbative approach to the study of the relativistic fluctuations of the geometry 
in a cold plasma. To reproduce the perturbative results it suffices to identify:
\begin{eqnarray}
N(\vec{x},\tau) &=& a(\tau) [1 +  \phi(\vec{x},\tau)], \qquad \gamma_{ij}(\vec{x},\tau) = 
a^2(\tau)[1 - 2 \psi(\vec{x},\tau)] \delta_{ij},
\nonumber\\
\rho_{\gamma}(\vec{x},\tau) &=& \overline{\rho}_{\gamma}(\tau) [ 1 + \delta_{\gamma}(\vec{x},\tau)].
\label{EL36}
\end{eqnarray}
Provided $\phi$, $\psi$ and $\delta_{\gamma}$ are perturbatively well 
defined, the choice of Eq. (\ref{EL36}) corresponds to the case of the conformally Newtonian gauge discussed, for related reasons, 
in appendix \ref{APPB}.  Using Eq. (\ref{EL36}), Eqs. (\ref{EL33}), (\ref{EL34}) and 
(\ref{EL35}) then become:
\begin{eqnarray}
\partial_{\tau} \delta_{\gamma} &=& 4 \partial_{\tau} \psi - \frac{4}{3} \vec{\nabla} \cdot \vec{v}_{\gamma},
\label{EL37}\\
\partial_{\tau} \vec{v}_{\gamma} &=& - \frac{1}{4} \vec{\nabla} \delta_{\gamma} - \vec{\nabla} \phi + 
a \Gamma_{\gamma\mathrm{e}} (\vec{v}_{\mathrm{b}} - \vec{v}_{\gamma}),
\label{EL38}\\
\partial_{\tau} \vec{v}_{\mathrm{b}} &=& - {\mathcal H} \vec{v}_{\mathrm{b}} + \frac{\vec{J} \times \vec{B}}{a^4 \rho_{\mathrm{b}}} - \vec{\nabla} \phi + \frac{4}{3} \frac{\rho_{\gamma}}{\rho_{\mathrm{b}}} a \Gamma_{\gamma \mathrm{e}}(\vec{v}_{\gamma}
-\vec{v}_{\mathrm{b}}),
\label{EL39}
\end{eqnarray}
which coincide with the system discussed in \cite{mg3}. For instance 
Eqs. (\ref{EL37}), (\ref{EL38}) and (\ref{EL39}) have been used to derive the appropriate 
initial conditions for magnetized CMB anisotropies in the tight-coupling approximation. 
In connection with Eq. (\ref{EL33}) we should mention a general feature of the interplay 
between the gradient expansion and the more conventional perturbative expansion leading to Eqs. (\ref{EL37}), 
(\ref{EL38}) and (\ref{EL39}): the term containing the velocity field in Eq. (\ref{EL33}) is of higher 
order in the gradient expansion but it is necessary to recover the correct linearized result. 
This occurrence simply signals that the gradient expansion and the standard perturbative 
expansion do not commute: the linearization of the equations written at the lowest order 
in the gradient expansion leads to a set of equations which is different from the one 
obtained by linearizing the exact equations. A useful compromise, as shown in Eq. (\ref{EL33}),
is to keep in the lowest order of the gradient expansion all those terms leading, after the linearization,
to the standard perturbative equations in a given gauge.

\renewcommand{\theequation}{4.\arabic{equation}}
\setcounter{equation}{0}
\section{Counting gradients in weakly coupled plasmas}
\label{sec4}
Let us consider, for sake of concreteness, typical length scales of the order of (or larger than) the Hubble radius \footnote{The definition of the Hubble 
radius $r_{\mathrm{H}}$ refers, strictly speaking, to the homogeneous and isotropic case and 
can be slightly ambiguous in the fully inhomogeneous situation which is the one 
treated in the present paper. For the moment, $r_{\mathrm{H}}$ must be considered as the inverse 
of the trace of the extrinsic curvature (possibly evaluated in a specific gauge). } (and hence larger than the 
Debye length) and typical time derivatives much smaller than the 
plasma frequency (see e.g. Eq. (\ref{OHM2})).  In weakly coupled plasmas the plasma parameter\footnote{The plasma parameter quantifies, by definition, the inverse of the 
number of particles present in the Debye sphere, i.e. the sphere whose radius is given by 
the Debye length.} $g_{\mathrm{plasma}}$ is, by definition very small
\begin{eqnarray}
&&g_{\mathrm{plasma}} = \frac{1}{V_{\mathrm{D}} n_{0} x_{\mathrm{e}}} = 24 e^{3}  \sqrt{\frac{\zeta(3)}{\pi}} 
\sqrt{x_{\mathrm{e}} \eta_{\mathrm{b}0}}
= 2.308 \times 10^{-7} \sqrt{x_{\mathrm{e}}} \biggl(\frac{h_{0}^2\Omega_{\mathrm{b0}}}{0.02273}\biggr)^{1/2},
\nonumber\\
&&V_{\mathrm{D}} = \frac{4}{3} \pi \lambda_{\mathrm{D}}^3,\qquad 
\lambda_{\mathrm{D}} =\sqrt{\frac{T}{8\pi e^2 n_{0} x_{\mathrm{e}}}},
\label{Red5}
\end{eqnarray}
where the reference temperature $T$ has been taken to coincide with the photon temperature (recall that the electron and proton temperatures coincide up to a small quantity which is the ratio of $r_{\mathrm{H}}^{-1}$ to the electron-photon rate \cite{mg1,mg2}). Given the smallness of $g_{\mathrm{plasma}}$, the electron-ion mean free path is much larger than 
the Debye scale while the corresponding collision frequency is much smaller than the plasma frequency, i.e. 
\begin{equation}
 \frac{\lambda_{\mathrm{D}}}{\lambda_{\mathrm{e\,i}}} = \frac{g_{\mathrm{plasma}}}{48\pi}
\ln{\Lambda_{\mathrm{C}}},\qquad \frac{\Gamma_{\mathrm{\mathrm{e\,i}}}}{\omega_{\mathrm{pe}}} = \frac{\ln{\Lambda_{\mathrm{C}}}}{24 \sqrt{2} \pi } g_{\mathrm{plasma}},
\label{Red4}
\end{equation}
where $\Lambda_{\mathrm{C}} = (18 \sqrt{2}/g_{\mathrm{plasma}})$ is the argument of the Coulomb logarithm.
Since $r_{\mathrm{H}}$ greatly exceeds $\lambda_{\mathrm{e i}}$ the space-time gradients of the system 
under consideration will be ordered as follows:
\begin{equation}
\nabla <  {\mathcal O}\biggl( \frac{1}{r_{\mathrm{H}}} \biggr) \ll {\mathcal O}\biggl( \frac{1}{\lambda_{\mathrm{e\, i}}}\biggr) \ll {\mathcal O}\biggl(\frac{1}{\lambda_{\mathrm{D}}}\biggr), \qquad 
\partial_{\tau} \ll {\mathcal O} (\omega_{\mathrm{p\,e}}),
\label{Red6}
\end{equation}
where $\nabla$ and $\partial_{\tau}$ denote, respectively, the magnitude 
of a typical spatial gradient and the magnitude of a typical time gradient.
Let us then consider Eq. (\ref{Red6})  in conjunction with Eq. (\ref{Red1}) and with Eq. 
(\ref{EL12}) whose explicit form, in terms of the total current, can also be written as:
\begin{equation}
(\vec{\partial} \times \vec{B})^{k} = 4 \pi J^{k}  + \partial_{\tau} E^{k}.
\label{Red7}
\end{equation}
Because of the hierarchies established in Eq. (\ref{Red6})  the time gradient of the total current 
must be negligible in comparison with the term containing the total rate. Similarly the total current must 
be much larger than the displacement current, i.e. 
\begin{equation}
\partial_{\tau} J^{k} \simeq {\mathcal O}\biggl( {\mathcal G}_{i}^{k} J^{i} \biggr) \ll N \Gamma J^{k}, \qquad 4\pi \vec{J}^{k} \gg \partial_{\tau} E^{k}.
\label{Red8}
\end{equation}
Equations (\ref{Red1}) and (\ref{Red6}) together with Eq. (\ref{Red8}) also imply the following 
pair of inequalities:
\begin{equation}
\partial_{\tau} J^{k} \ll \omega_{\mathrm{pe}}^2 \partial_{\tau} E^{k},\qquad 
\partial_{\tau}^{2} J^{k} \ll \omega_{\mathrm{pe}}^2 J^{k}.
\label{Red9}
\end{equation}
But then Eq. (\ref{Red9}) means that the Ohm equation of Eqs. (\ref{OHM3}) and (\ref{Red1}) 
reduces to the following Ohm law: 
\begin{equation}
J^{k}= \sigma\biggl[ E^{k} + 
(\vec{v}_{\mathrm{b}} \times \vec{B})^k\biggr], \qquad \sigma= \frac{\omega_{\mathrm{pe}}^2}{4\pi N \Gamma_{\mathrm{i e}}},
\label{OHM4}
\end{equation}
where the conductivity $\sigma$ is not bound to be homogeneous 
as in the case of previous treatments \cite{mg3}.  Equation (\ref{OHM4}) generalizes 
to the fully inhomogeneous situation the standard result of the conformally flat limit already mentioned in Eq. (\ref{EL14}) (see, e.g. \cite{mg2}, third reference). Note that, in the latter limit, the spatial derivatives of the extrinsic 
curvature and of the determinant of the metric are all vanishing; moreover 
the total current, the electric and magnetic fields, the conductivity are all rescaled through different powers of the scale factor with respect to their flat-space values. 
Consider now Eq. (\ref{OHM4}) written in its explicit form. Because 
of Eqs. (\ref{Red7}), (\ref{Red8}) and (\ref{Red9}) the Ohmic electric field can be expressed as:
\begin{equation}
E^{i} = \frac{1}{4\pi \sigma} \partial_{j} \biggl[ N \gamma^{i k} \, \gamma^{j n} \, \eta_{m n k} \, B^{m} \biggr]  - \frac{1}{N} \gamma_{f q} \, \gamma_{g p} \, \eta^{q p i} \,v_{\mathrm{b}}^{f} \, B^{g},
\label{Red10}
\end{equation}
where the second equation reported in Eq. (\ref{EL10}) has been used to simplify the obtained result.
Equation (\ref{Red10}) still implies, for weakly coupled plasmas, that 
in the drift approximation the electric fields depend upon the gradients 
of the magnetic field in the baryon rest frame. This means 
that the electric fields are higher order in the gradients. From Eq. (\ref{Red10}), 
the generalized magnetic diffusivity equation can be derived and the result is 
\begin{equation}
\partial_{\tau} B^{h} + \frac{1}{4\pi} \partial_{r} \biggl\{ \frac{N}{\sigma} \gamma^{ h s}
\gamma^{r u} \eta_{s u i} \partial_{j} \biggl[ N\,\gamma^{ik}\,\gamma^{j n}\,\eta_{n m k} 
\,B^{m} \biggr]\biggr\} = \partial_{r} \biggl[ \gamma^{ h s}\,\gamma^{r u}\, \eta_{s u i}\,
\gamma_{f q}\,\gamma_{g p}\,v_{\mathrm{b}}^{f}\,B^{g}\,\eta^{q p i}\biggr],
\label{Red11}
\end{equation}
where, for immediate convenience, we preferred to avoid the shorthand notations employed before.
The analog equation for the electric field is instead 
\begin{equation}
\partial_{\tau} E^{i} = - \partial_{\tau} \biggl( \frac{\eta^{q p i}}{N} \, \gamma_{f q} \, 
\gamma_{g p} \, v_{\mathrm{b}}^{f} \, B^{g} \biggr)
+ \frac{1}{4 \pi \sigma} \partial_{j} \biggl[ \partial_{\tau} \biggl( N\, \gamma^{i k} \, \gamma^{j n} \, \eta_{m n k} B^{m} \biggr)\biggr].
\label{Red12}
\end{equation}
In the conformally flat limit mentioned in Eq. (\ref{EL14}), Eqs. (\ref{Red11}) and (\ref{Red12}) coincide, respectively, 
with their flat space counterparts, namely:
\begin{eqnarray} 
&&\frac{\partial \vec{B}}{\partial \tau} = \vec{\nabla} \times (\vec{v}_{\mathrm{b}} \times \vec{B})+ \frac{1}{4\pi \sigma} \nabla^2 \vec{B},
\label{Red13}\\
&& \frac{\partial \vec{E}}{\partial \tau} = -\frac{\partial}{\partial\tau} (\vec{v}_{\mathrm{b}} \times \vec{B})
+ \frac{1}{4\pi \sigma} \nabla^2 \vec{E},
\label{Red14}
\end{eqnarray}
where, according to Eq. (\ref{EL9}),  $\vec{B} = a^2 \, \vec{{\mathcal B}}$ 
and $\vec{E} = a^2 \vec{{\mathcal E}}$. 

Let us now rewrite Eqs. (\ref{EE3}), (\ref{EE4}) and 
(\ref{EE5}) in a more explicit form which will turn out to be useful in the forthcoming sections: 
\begin{eqnarray}
&&\partial_{\tau} K - N {\mathrm Tr} K^2 + \nabla^2 N = N \ell^2_{\mathrm{P}}\biggl[ 
\frac{3 p + \rho}{2} + (p + \rho) s^2(y) + \rho_{\mathrm{B}} + \rho_{\mathrm{E}}\biggr],
\label{AN1}\\
&& \nabla_{i} K - \nabla_{k} K^{k}_{i} = \ell_{\mathrm{P}}^2 \biggl[ (p + \rho) 
\hat{v}_{i} s(y) c(y) - \frac{1}{4 \pi \gamma} \eta^{m n p} \gamma_{i m} \gamma_{j n} \gamma_{p k}
B^{k} E^{j} \biggr], 
\label{AN2}\\
&& \partial_{\tau} K_{i}^{j} - N K K_{i}^{j} - N r_{i}^{j} + \nabla_{i} \nabla^{j} N = 
N \ell_{\mathrm{P}}^2 \biggl[ \frac{p - \rho}{2} \delta_{i}^{j} - ( p + \rho) \, s^2(y) \,\hat{v}_{i} \hat{v}^{j}
\nonumber\\
 &-& ( p_{\mathrm{B}} + p_{\mathrm{E}}) \delta_{i}^{j} + \tilde{\Pi}_{i}^{j} + \Pi_{i}^{j}(E) + \Pi_{i}^{j}(B)\biggr],
\label{AN3}
\end{eqnarray}
where the notations of appendix \ref{APPA} have been used (see, in particular, Eqs. (\ref{T14})--(\ref{T15}) and (\ref{T24})--(\ref{T26})).  To zeroth order in the gradient expansion the terms containing $s(y)$ are subleading since 
$y = v/N$ and $v^2 = \gamma_{i j} v^{i} v^{j}$. 
By combining Eq. (\ref{AN1}) and the trace of Eq. (\ref{AN3})
the terms $\partial_{\tau} K$ can be eliminated and the resulting expression is the Hamiltonian constraint
\begin{equation}
K^2 -\mathrm{Tr} K^2 + r= \ell_{\mathrm{P}}^2 [ 2 ( p + \rho) s^2(y) + 2 \rho + 
(\rho_{\mathrm{E}} + 3 p_{\mathrm{E}}) + (\rho_{\mathrm{B}}+ 3 p_{\mathrm{B}})]
\label{AN4}
\end{equation}
coinciding, as expected, with the expression already obtained in Eq. (\ref{EE6}) in the light of the values of $p_{\mathrm{B}}$ and $p_{\mathrm{E}}$.  We recall that the velocity field, the energy density $\rho$, the pressure $p$ are global quantities given by the sum over the individual species:
\begin{equation}
 p = \sum_{\mathrm{a}} p_{\mathrm{a}}, \qquad 
\rho = \sum_{\mathrm{a}} \rho_{\mathrm{a}},\qquad ( p + \rho) v^{k} = \sum_{\mathrm{a}} (p_{\mathrm{a}} + \rho_{\mathrm{a}}) v_{\mathrm{a}}^{k},
\end{equation}
and obeying conservation equations which can be obtained, species by species, using the results of Eqs. (\ref{T9}) and (\ref{T10}) reported in appendix \ref{APPA}:
\begin{eqnarray}
&& \partial_{\tau} \rho + N \partial_{k} \biggl[\frac{(p + \rho)}{N} v^{k}\biggr] - N K ( p + \rho) =0,
\label{AN5}\\
&& \partial_{\tau} [ ( p + \rho) v^{i}] + N^2 \partial_{k} ( p \gamma^{i k}) + N^2 \partial_{k} 
\Pi^{k i} + (p + \rho) N\partial^{i} N - 2 N K_{j}^{i} (p + \rho) v^{j} 
\nonumber\\ 
&& - N K (p + \rho) v^{i} 
=0.
\label{AN6}
\end{eqnarray}
As already mentioned after Eq. (\ref{Red10}),  
the baryon rest frame is particularly useful for the treatment of the finite 
conductivity effects. In a perfectly conducting medium (i.e. $\sigma(\vec{x},\tau) \to \infty$) the Ohmic electric field is perfectly screened. Owing to this occurrence, it is customary to define the 
plasma frame where the electric fields are set to zero. In the usual perturbative 
expansion defined, for instance, in appendix \ref{APPB}, the plasma frame coincides with the baryon rest frame. 
In the fully nonlinear case, however, the two concepts do not necessarily coincide. 
It is finally appropriate to mention that there are also nonlinearities 
associated with the Ohm law itself (see Eq. (\ref{OHM3})) and an example along this direction is 
the so-called nonlinear Hall effect. This effect comes by retaining the terms $\vec{J}\times \vec{B}$ in Eq. (\ref{OHM3}). 
The Hall term leads, in the magnetic diffusivity equation (\ref{Red13}), to a term of the type $\vec{\nabla} \times [\vec{B} \times (\vec{\nabla} \times \vec{B})]/(e n_{0})$. The nonlinear Hall term can partially balance or even become greater than the dynamo term, under certain conditions \cite{KLE}. The present framework paves the way for the consistent treatment 
of the  gravitating counterpart of the nonlinear effects typical of cold plasmas (see, for instance, \cite{biskamp}).

\renewcommand{\theequation}{5.\arabic{equation}}
\setcounter{equation}{0}
\section{Anti-Newtonian drift approximation}
\label{sec5}
Consider, for simplicity, the case where $N=N(\tau)$; then $\gamma_{ij}(\vec{x},\tau)$ corresponds to $6$ unknown functions and all the terms containing at least one spatial gradient of $N$ vanish exactly. The spatial curvature is neglected 
since it contains two spatial gradients of $\gamma_{ij}$. Bearing in mind the definition of Eq. (\ref{EE8}), Eqs. (\ref{AN1}) and (\ref{AN3}) can be written as 
\begin{eqnarray}
&& \partial_{\tau} K = N \mathrm{Tr} K^2 + \frac{N \ell_{\mathrm{P}}^2}{2} \biggl[ ( 3 \overline{p} + \overline{\rho}) - 2 \rho_{\Lambda} + 2 \rho_{\mathrm{B}} \biggr],
\label{AN2aa}\\
&& \partial_{\tau} K = N K^2 + \frac{3 \ell_{\mathrm{P}}^2 N}{2} [ \overline{p} - \overline{\rho} - 2 \rho_{\Lambda} - 2 p_{\mathrm{B}} ],
\label{AN2bb}\\
&& \partial_{\tau} \overline{K}_{i}^{j}  = N K \overline{K}_{i}^{j} + \ell_{\mathrm{P}}^2 \,N\, \Pi_{i}^{j},
\label{AN2cc}\\
&& \partial_{\tau} \overline{\rho} = N K (\overline{p} + \overline{\rho}), \qquad \partial_{\tau} B^{i}= 0.
\label{AN2dc}
\end{eqnarray}
Concerning Eqs. (\ref{AN2aa})--(\ref{AN2dc}) few comments are in order. The Ohmic electric fields have been neglected 
since they are of higher order and, for sake of simplicity, $\Pi_{i}^{j}$ denotes the {\em total} anisotropic stress. Note that, in Eqs. (\ref{AN2aa})--(\ref{AN2dc}), the total pressure and the total energy density 
have been separated, respectively, as 
$p = \overline{p} - \rho_{\Lambda}$ and $\rho = \overline{\rho} + \rho_{\Lambda}$ where $\rho_{\Lambda}$ parametrizes 
the dark energy density contribution;
 $\overline{p}$ and $\overline{\rho}$ are the pressure and energy density of an ordinary fluid 
 characterized by a (possibly inhomogeneous) barotropic index $w = \overline{p}/\overline{\rho}$. 
The approximations leading to Eqs. (\ref{AN2aa})--(\ref{AN2dc})  define the anti-Newtonian limit. 
Equations (\ref{AN2aa}) and (\ref{AN2bb}) can also be written in more explicit terms by introducing 
the barotropic index and by recalling that, as in Eq. (\ref{EE8}), $\overline{K}_{i}^{j} = K_{i}^{j} - (K/3) \delta_{i}^{j}$:
\begin{eqnarray}
&& \partial_{\tau} K = N K^2 + \frac{3 N \ell_{\mathrm{P}}^2}{2} [ ( w -1) \overline{\rho} - 2\rho_{\Lambda} - 2 p_{\mathrm{B}}],
\label{AN3aa}\\
&& \partial_{\tau} K = \frac{N}{3} K^2 + N \mathrm{Tr} \overline{K}^2 + \frac{N \ell^2_{\mathrm{P}}}{2}[ ( 3 w + 1) \overline{\rho} - 2 \rho_{\Lambda} + 2 \rho_{\mathrm{B}}].
\label{AN3bb}
\end{eqnarray}
By eliminating $\overline{\rho}$ between Eqs. (\ref{AN3aa}) and (\ref{AN3bb}) and by defining  $ d t = N(\tau) d\tau$, the resulting equation is given by
\begin{equation}
\partial_{t} K - \frac{w+1}{2} K^2 + \frac{3( w-1)}{4} \mathrm{Tr} \overline{K}^2 = 
- \frac{3\ell_{\mathrm{P}}^2}{2} [( w + 1) \rho_{\Lambda}+ \frac{( 3w -1)}{3} \rho_{\mathrm{B}}].
\label{AN4aa}
\end{equation}
The explicit solution of Eq. (\ref{AN2cc}), $\overline{K}_{i}^{j}(\vec{x},t)$ implies that  
\begin{equation}
\overline{K}_{i}^{j}(\vec{x},t) = \frac{\lambda_{i}^{j}(\vec{x})}{\sqrt{\gamma(\vec{x},t}} + \ell_{\mathrm{P}}^2 \frac{\sigma_{i}^{j}(\vec{x},t)}{\sqrt{\gamma(\vec{x},t)}},\qquad 
\sigma_{i}^{j}(\vec{x},t) = \int^{t} \sqrt{\gamma(\vec{x}, t')}\, \Pi_{i}^{j}(\vec{x},t') \, d t'.
\label{AN5aa}
\end{equation}
By now introducing the rescaled variable 
${\mathcal M}(\vec{x},t) = [\gamma(\vec{x},t)]^{(w +1)/4}$, and by recalling that $\rho_{\mathrm{B}}(\vec{x}, t) = \rho_{\mathrm{B}}(\vec{x}) \gamma^{-2/3}$, Eq. (\ref{AN4aa}) 
\begin{eqnarray}
\partial_{t}^2 {\mathcal M} &=& \frac{3}{8} (w^2 -1) {\mathcal M}^{\frac{w - 3}{w +1}}[ \mathrm{Tr} \lambda^2 
+ \ell_{\mathrm{P}}^4 \mathrm{Tr} \sigma^2 + 2 \ell_{\mathrm{P}}^2 {\mathrm Tr} \lambda \, \sigma] 
\nonumber\\
&+& \frac{ 3 ( w + 1)^2\ell_{\mathrm{P}}^2}{4}  \rho_{\Lambda}(\vec{x}) \biggl\{ 1 + \frac{3 w -1}{3(w + 1)}  \Lambda_{\mathrm{B}}(\vec{x}) {\mathcal M}^{- 8/[3( w +1)]}\biggr\}\, {\mathcal M},
\label{AN7aa}
\end{eqnarray}
where $\Lambda_{\mathrm{B}}(\vec{x}) = \rho_{\mathrm{B}}(\vec{x})/\rho_{\Lambda}(\vec{x})$.
Defining the following auxiliary coefficients 
\begin{equation}
{\mathcal C}(\vec{x},t) =  \frac{3}{8} (w^2 -1) [ \mathrm{Tr} \lambda^2 
+ \ell_{\mathrm{P}}^4 \mathrm{Tr} \sigma^2 + 2 \ell_{\mathrm{P}}^2 {\mathrm Tr} \lambda \, \sigma], \qquad 
{\mathcal D}(\vec{x}) =  \frac{ 3 ( w + 1)^2\ell_{\mathrm{P}}^2}{4}  \rho_{\Lambda}(\vec{x}),
\label{COEF1}
\end{equation}
Eq. (\ref{AN7aa}) becomes 
\begin{equation}
\partial_{t}^2 {\mathcal M}= {\mathcal C}(\vec{x},t) {\mathcal M}^{\frac{w - 3}{w +1}} + {\mathcal D}(\vec{x})  \biggl\{ 1 + \frac{3 w -1}{3(w + 1)}  \Lambda_{\mathrm{B}}(\vec{x}) {\mathcal M}^{- 8/[3( w +1)]}\biggr\}\, {\mathcal M}.
\label{COEF2}
\end{equation}
The general form of Eq. (\ref{COEF2}) does not have explicit analytic solutions. In various limits it is possible to 
integrate it once with respect to $t$ in terms of space-dependent integration constants. The 
result, however, cannot be further integrated (or inverted) except that in few special cases. Eq. (\ref{COEF2}) 
can certainly be integrated numerically but this study is beyond the scopes of the present discussion.

It is useful to look at Eq. (\ref{COEF2}) by bearing in mind the usual assumptions of 
the separate Universe picture stipulating  that any portion of the Universe that is larger than the Hubble radius $r_{\mathrm{H}}$ but smaller than the physical wavelength 
on the perturbation  will look like a separate unperturbed Universe.  
Such a framework is justified in the case $\overline{K}_{i}^{j} =0$.  In the latter case, the
extrinsic curvature can be really thought as the inhomogeneous generalization 
of the Hubble parameter. Assuming, for consistency with this hypothesis, that large-scale 
electromagnetic fields are absent, Eqs. (\ref{AN2aa})--(\ref{AN2cc})
formally coincide with the Friedmann-Lema\^itre equations.  
Consider next a slightly more complicated situation, namely the one where 
$\rho_{\Lambda}=0$ but $\overline{K}_{i}^{j} \neq 0$. In this case Eq. (\ref{COEF2}) can be explicitly integrated 
and ${\mathcal M}$ can be determined. Suppose, for sake of simplicity, that $w=0$ and that $\Lambda_{\mathrm{B}}(\vec{x})=0$. In this case Eq. (\ref{COEF2}) can be solved exactly: ${\mathcal C}(\vec{x},t)$ becomes independent of $t$ since 
$\sigma_{i}^{j}(\vec{x},t) \to 0$. The determinant of the metric becomes, in the cosmic time coordinate $t$:
\begin{equation}
\gamma(\vec{x},t) = \biggl\{ H_{\mathrm{i}}^2\biggl[ t - t_{\mathrm{i}}(\vec{x})\biggr]^2 - \frac{3}{8} \frac{\mathrm{Tr} \lambda^2}{H_{\mathrm{i}}^2}
\biggr\}^2,
\label{AN11aa}
\end{equation}
where $H_{\mathrm{i}}$ is a space-time constant while, as already mentioned, $\mathrm{Tr}\lambda^2$ 
can have an arbitrary spatial dependence.  
From Eq. (\ref{AN11aa}) we can argue that even if $\overline{K}_{i}^{j} \neq 0$, $\mathrm{Tr}\lambda^2$ 
affects the final solution in such a way that it can be reabsorbed in the definition of the initial time coordinate or, put it in different terms, $\overline{K}_{i}^{j} \neq 0$ but goes quickly to zero if the Universe expands since  $\overline{K}_{i}^{j} = \lambda_{i}^{j}/\sqrt{\gamma}$.  If the magnetic anisotropic stress is taken into account the full system must be solved consistently 
and anti-Newtonian solutions do not seem sufficient even if general solutions of Eq. (\ref{COEF2}) would be 
available. Therefore it seems useful to explore a slightly different strategy and solve the system in the quasi-isotropic limit where the contribution of the intrinsic curvature and of the magnetic anisotropic stress can be explicitly taken into account order by order.

\renewcommand{\theequation}{6.\arabic{equation}}
\setcounter{equation}{0}
\section{Quasi-isotropic MHD solution}
\label{sec6}
In the MHD description, the fields $E^{i}$, $B^{i}$ and $J^{i}$ defined in Eqs. (\ref{EL9}) and (\ref{EL29}) are solenoidal, i.e. 
\begin{equation}
\partial_{i} E^{i} =0, \qquad \partial_{i} B^{i} =0,\qquad \partial_{i} J^{i} =0,
\label{SOL1}
\end{equation}
so that flat space MHD is recovered in the conformally flat limit of Eq. (\ref{EL14}).  
Let us consider, as first example of the present section, the following parametrization of the spatial geometry 
\begin{equation}
\gamma_{ij}(\vec{x}, \tau) = a^2(\tau) e^{ - 2 \Psi(\vec{x},\tau)}  [\delta_{ij} + 2 h_{ij}(\vec{x},\tau)],
\qquad h_{i}^{i}=0,\qquad N(\tau) = a(\tau),
\label{PART1}
\end{equation}
where $h_{ij}$ accounts for terms containing more than one spatial gradient. Using Eq. (\ref{PART1}) into 
Eq. (\ref{ADM1a}) the explicit form of the extrinsic curvature becomes
\begin{equation}
K_{i}^{j} = \biggl[ \biggl( \frac{\partial_{\tau} \Psi}{a} - \frac{{\mathcal H}}{a} \biggr) 
\delta_{i}^{j} - \frac{\partial_{\tau} h_{i}^{j}}{a} \biggr].
\label{PART2}
\end{equation}
The conformally flat limit of Eqs. (\ref{PART1}) and 
(\ref{PART2}) corresponds to $\Psi \to 0$ and $h_{ij} \to 0$; following the same logic 
the total fluid pressure and the total energy density can be separated, respectively, as
\begin{equation}
p(\vec{x},\tau) = p^{(0)}(\tau)  + p^{(1)}(\vec{x},\tau), \qquad \rho(\vec{x},\tau) = \rho^{(0)}(\tau) + \rho^{(1)}(\vec{x},\tau),
\label{PAR1A}
\end{equation}
where $p^{(1)}(\vec{x},\tau)$ and $\rho^{(1)}(\vec{x},\tau)$ vanish in the conformally flat limit.
Inserting Eqs. (\ref{PART1})--(\ref{PART2}) into Eqs. (\ref{Red10})--(\ref{Red11})
the Ohmic electric field and the magnetic diffusivity equation are, respectively\footnote{The vectors appearing 
hereunder and in the remaining part of the present section are the standard 
three-dimensional vectors. Similarly the vector products are standard vector products.},
\begin{eqnarray}
&&\vec{E} = \frac{1}{4 \pi \sigma} \vec{\nabla}\times\biggl( e^{ \Psi}  \vec{B}\biggr)
- e^{- \Psi} \vec{v}_{\mathrm{b}} \times \vec{B},
\label{PART3}\\
&& \partial_{\tau} \vec{B} + \frac{1}{4 \pi \sigma} \vec{\nabla} \times \biggl[ e^{\Psi} \vec{\nabla}
\times\biggl( e^{\Psi} \vec{B} \biggr)\biggr] = \vec{\nabla} \times ( \vec{v}_{\mathrm{b}} \times \vec{B}).
\label{PART4}
\end{eqnarray}
The terms containing $h_{ij}$ have been neglected since they are of higher order.
Inserting Eqs. (\ref{PART1}) and (\ref{PART2}) into Eq. (\ref{EE3}) the following 
relation can be obtained
\begin{eqnarray}
 \partial_{\tau}^2 \Psi + {\mathcal H} \partial_{\tau} \Psi - \partial_{\tau} {\mathcal H} &=& a^2 \ell_{\mathrm{P}}^2 \biggl\{ \frac{\rho + 3 p}{6} 
\nonumber\\
&+& \frac{( p + \rho)}{3} a^2 e^{ - 2 \Psi} s^2(y) +
\frac{ e^{4 \Psi}}{24 \pi a^4} \biggl[ E^2 + \tilde{E}^2 + B^2 + \tilde{B}^2\biggr]\biggr\},
\label{PART5}
\end{eqnarray}
where the following notations have been adopted:
\begin{eqnarray}
&& E^2 = \delta_{ij} E^{i} E^{j},\qquad \tilde{E}^2 = 2 h_{i j} E^{i} E^{j}, \qquad 
B^2 = \delta_{ij} B^{i} B^{j}, \qquad \tilde{B}^2 = 2 h_{ij} B^{i} B^{j} 
\nonumber\\
&& u^2 = \gamma_{i j} u^{i} u^{j} = a^2 e^{- 2 \Psi} [ s^2(y) + 2 c^2(y) h_{ij} \hat{v}^{i} 
\hat{v}^{j}],\qquad y = \frac{v}{a},\qquad v^2 = \delta_{ij} v^{i} v^{j}.
\label{PART6}
\end{eqnarray}
Note that   $\tilde{E}^2 \ll E^2$ and $\tilde{B}^2 \ll B^2$ since $\tilde{E}$ and $\tilde{B}$
contain $h_{ij}$ and are therefore of higher order in the expansion. Furthermore, in the resistive MHD limit (i.e. 
$\sigma \gg 1$)
$\rho_{\mathrm{E}} \ll \rho_{\mathrm{B}}$ since 
\begin{eqnarray}
\rho_{\mathrm{E}} &=& \frac{1}{8 \pi a^4} \biggl\{ \frac{e^{6 \Psi}}{16 \pi^2 \sigma^2} \biggl[ \bigl| \vec{\nabla}\times \vec{B} \bigr|^2 + 
\bigl| \vec{\nabla}\Psi \times \vec{B} \bigr|^2 + 2 (\vec{\nabla} \Psi \times \vec{B}) \cdot (\vec{\nabla} \times \vec{B})\biggr]
\nonumber\\
&-& \frac{e^{4 \Psi}}{2 \pi \sigma} \biggl[( \vec{v}_{\mathrm{b}} \times \vec{B})\cdot (\vec{\nabla}\times \vec{B}) + ( \vec{v}_{\mathrm{b}} \times 
\vec{B}) \cdot (\vec{\nabla} \Psi \times \vec{B})\biggr]  + e^{ 2 \Psi} \bigl| \vec{v}_{\mathrm{b}} \times \vec{B} \bigr|^2\biggr\}.
\label{PART9}
\end{eqnarray}
Equation (\ref{PART9}) shows, in practice, that at the nonlinear level the baryon rest frame and the 
plasma frame are two complementary concepts. While in the baryon rest frame the electric energy density 
vanishes as $\sigma^{-2}$ for large conductivity, if the conductivity is not large the contributions 
of the baryon velocity and of the gradients of the geometry are of higher order in the spatial gradients 
in comparison with the magnetic contributions. 
The resistive MHD limit implies, as usual, the largeness of the magnetic Reynolds 
number defined as 
\begin{equation}
R_{\mathrm{m}} = \frac{u}{\eta \nabla} \gg 1, \qquad \eta = \frac{1}{4\pi\sigma},
\label{REY}
\end{equation}
where $\eta$ is the magnetic diffusivity, $\nabla^{-1}$ denotes the typical 
length-scale associated with spatial gradients and $u$ is the modulus of the typical velocity 
of the plasma element.  
The velocity field itself is of higher order in comparison with $\rho$, $p$, $\rho_{\mathrm{B}}$ since, from the 
momentum constraint of Eq. (\ref{AN2}) we can deduce 
\begin{equation}
u^{0} u_{i}= \frac{1}{a^2 \ell_{\mathrm{P}}^2 ( p^{(0)} + \rho^{(0)})}\biggl\{ 2 \partial_{i} \partial_{\tau} \Psi +
\partial_{k} \partial_{\tau} h^{k}_{i} + \frac{\ell_{\mathrm{P}}^2 \,e^{ 3 \Psi}}{4 \pi a^2} (\vec{E} \times \vec{B})_{i} \biggr\},
\label{PART7}
\end{equation}
where, following the notations of Eqs. (\ref{EL7a}),  and to lowest order 
in the gradient expansion, the expression $u^{0} u_{i}$ can also be written as $u^{0} u_{i} = e^{- 2 \Psi} v_{i} c^2(y)$. The term $\vec{E}\times\vec{B}$  can be written by using Eq. (\ref{PART3}):
\begin{equation}
\vec{E} \times \vec{B} = e^{ \Psi} \biggl[ \frac{(\vec{\nabla} \times \vec{B}) \times \vec{B} + (\vec{\nabla} \Psi \times \vec{B}) \times \vec{B}}{4 \pi \sigma}\biggr] - e^{- \Psi} (\vec{v}_{\mathrm{b}} \times \vec{B}) \times \vec{B}.
\label{PART8}
\end{equation}
Neglecting the Ohmic electric fields, Eq. (\ref{AN3}) leads to the evolution equation for $h_{i}^{j}$:
\begin{eqnarray}
\partial_{\tau}^2h_{i}^{j} + 2 {\mathcal H} \partial_{\tau} h_{i}^{j}  &=& -e^{ 2 \Psi} \biggl\{ \biggl[ \partial_{i} \partial^{j} \Psi - \frac{1}{3} \nabla^2 
\Psi \delta_{i}^{j} \biggr] + \partial_{i} \Psi \partial^{j} \Psi - \frac{\delta_{i}^{j}}{3} (\nabla \Psi)^2 \biggr\}
\nonumber\\
&-&  a^2 \ell_{\mathrm{P}}^2 \biggl\{ - (p + \rho) \biggl(u_{i} u^{j} - \frac{u^2}{3} \delta_{i}^{j} \biggr) 
+ \frac{e^{4 \Psi}}{4 \pi a^4} \biggl(B_{i} B^{j} - \frac{B^2}{3} \delta_{i}^{j} \biggr)\biggr\},
\label{PART10}
\end{eqnarray}
where, using Eq. (\ref{PART1}) into Eq. (\ref{ADM1b}) and keeping only the leading contribution in the spatial 
gradients,  $r_{ij}(\vec{x},\tau)$ becomes, in explicit terms,
\begin{equation}
r_{ij}(\vec{x},\tau) = [ \nabla^2 \Psi - (\nabla \Psi)^2 ] \delta_{ij} + \partial_{i} \partial_{j} \Psi + \partial_{i} \Psi \partial_{j} \Psi,
\label{PART10a}
\end{equation}
where $\nabla^2 \Psi = \delta^{ij} \partial_{i} \partial_{j} \Psi$ is the flat-space Laplacian and, similarly, 
 $(\nabla\Psi)^2 = \delta^{ij} \partial_{i} \Psi\partial_{j} \Psi$.
 Finally,  the trace of Eq. (\ref{AN3}) leads, in the case of Eq. (\ref{PART1}), to the following condition 
 \begin{eqnarray}
&& \partial_{\tau}^2 \Psi + 5 {\mathcal H} \partial_{\tau} \Psi - (2 {\mathcal H}^2 + \partial_{\tau} {\mathcal H}) = \frac{2}{3} 
e^{2 \Psi}\biggl[ 2 \nabla^2 \Psi - (\nabla\Psi)^2\biggr]
\nonumber\\
&&+ a^2 \ell_{\mathrm{P}}^2 \biggl\{ \frac{p - \rho}{2} - \frac{p + \rho}{3} u^2 - 
\frac{ e^{4 \Psi}}{24 \pi a^4} \biggl[ B^2 + \tilde{B}^2\biggr]\biggr\}.
\label{PART11}
\end{eqnarray}
Using the parametrization of Eq. (\ref{PAR1A}), Eqs. (\ref{PART5}) and (\ref{PART11}) 
imply the following relations:
\begin{eqnarray}
&& \partial_{\tau}^2 \Psi + {\mathcal H} \partial_{\tau} \Psi = \frac{a^2 \ell_{\mathrm{P}}^2}{6} ( 3 p^{(1)} + \rho^{(1)}) + 
\frac{a^2 \ell_{\mathrm{P}}^2}{3} \rho_{\mathrm{B}},
\label{PART13}\\
&& \partial_{\tau}^2 \Psi + 5 {\mathcal H} \partial_{\tau} \Psi = \frac{2 e^{2\Psi}}{3} [ 2 \nabla^2 \Psi - (\nabla\Psi)^2] 
+ a^2 \ell_{\mathrm{P}}^2 \biggl[ \frac{p^{(1)} - \rho^{(1)}}{2} - p_{\mathrm{B}} \biggr]. 
\label{PART14}\\
&& 3 {\mathcal H}^2 = \ell_{\mathrm{P}}^2 a^2 \rho^{(0)}, \qquad 2({\mathcal H}^2 - \partial_{\tau} {\mathcal H}) = \ell_{\mathrm{P}}^2 
a^2 (\rho^{(0)} + p^{(0)}).
\label{PART14A}
\end{eqnarray}
Equations (\ref{PART10}), (\ref{PART13}) and (\ref{PART14}) can be directly solved and the result, in the plasma 
frame, reads:
\begin{eqnarray}
\Psi(\vec{x},\tau) &=&  q(\vec{x})+ \frac{2 \, a^{3 w + 1} \, e^{ 2 q(\vec{x})}}{3 ( 3 w + 5) H_{\mathrm{i}}^2} \biggl\{ \nabla^2 q(\vec{x}) - 
\frac{[\nabla q(\vec{x})]^2}{2}\biggr\} - \frac{\ell_{\mathrm{P}}^2\, a^{ 3 w -1}\, e^{4 q(\vec{x})}}{ 60 \pi ( 3 w +1) H_{\mathrm{i}}^2} B^2(\vec{x})
\label{PART15}\\
h_{i}^{j}(\vec{x},\tau) &=& - \frac{2 \, a^{ 3 w +1}\, e^{ 2 q(\vec{x})}}{( 3 w +1 ) ( 3 w + 5) H_{\mathrm{i}}^2}\biggl\{ \biggl[ \partial_{i} 
\partial^{j} q(\vec{x}) - \frac{\nabla^2 q(\vec{x})}{3} \delta_{i}^{j} \biggr] + \partial_{i} q(\vec{x}) \partial^{j} q(\vec{x}) - \frac{[\nabla q(\vec{x})]^2}{3} \delta_{i}^{j}\biggr\}
\nonumber\\
&-& \frac{ \ell_{\mathrm{P}}^2\, a^{3 w -1}\, e^{ 4 q(\vec{x})}}{ 2 \pi (3 w -1) ( 3 w+1)}  \biggl[ B_{i}(\vec{x})B^{j}(\vec{x}) - \frac{B^2(\vec{x})}{3} 
\delta_{i}^{j}\biggr],
\label{PART16}
\end{eqnarray}
where $q = q(\vec{x})$ is a generic function of the spatial coordinates encoding the dependence of the 
large-scale inhomogeneities. Note also 
that we used $H_{\mathrm{i}} = {\mathcal H}_{\mathrm{i}}/a_{\mathrm{i}}$ at the initial 
reference time $\tau_{\mathrm{i}}$. Equations (\ref{PART15}) and (\ref{PART16}) 
show that the relative importance of the spatial gradients does depend upon the barotropic index.

The results obtained in the particular case of Eq. (\ref{PART1}) can also 
be deduced in a  more general parametrization, namely:
\begin{equation}
\gamma_{ik} = a^2(\tau) [ \alpha_{ik}(\vec{x}) + \beta_{ik}(\vec{x},\tau)],\qquad
\gamma^{kj} = \frac{1}{a^2(\tau)} [ \alpha^{kj}(\vec{x}) - \beta^{kj} (\vec{x},\tau)],\qquad N(\tau) = a(\tau),
\label{exp1}
\end{equation}
where $\beta_{ij}(\vec{x},\tau)$ contains the contribution of the gradients while $\alpha_{ij}(\vec{x})$ 
is fully inhomogeneous but does not contain any gradient.  The conformally flat 
limit of Eq. (\ref{exp1}) does correspond to $\alpha_{ij} \to \delta_{ij}$ and $\beta_{ij} \to 0$. The 
indices of $\beta_{ij}$ are raised and lowered by using $\alpha_{ij}$.
According to Eq. (\ref{ADM1a}), the extrinsic curvature and its contractions become:
\begin{equation}
K_{i}^{j} = - \frac{1}{a}\biggl( {\mathcal H} \delta_{i}^{j} + 
\frac{\partial_{\tau}{\beta}_{i}^{j}}{2} \biggr),
\qquad K = - \frac{1}{a}\biggl( 3 {\mathcal H} + 
\frac{1}{2}\partial_{\tau}\beta\biggr),
\qquad {\rm Tr} K^2 = \frac{1}{a^2} \biggl(3 {\mathcal H}^2 
+ {\mathcal H} \partial_{\tau} \beta\biggr),
\label{exp2}
\end{equation}
where, as in the previous case, $ {\mathcal H}= \partial_{\tau} \ln{a}$.
From the momentum constraint it also follows that 
\begin{equation}
\nabla_{k} \partial_{\tau} \beta_{i}^{k} - \nabla_{i} \partial_{\tau} \beta = 2 a \ell_{\mathrm{P}}^2  \hat{v}_{i} c(y) s(y) (p+ \rho).
\label{momex}
\end{equation}
The explicit form of the momentum constraint suggests to 
look for the solution in a separable form, namely, 
\begin{equation} 
\beta_{i}^{j}(\vec{x},\tau) = g(\tau) \mu_{i}^{j}(\vec{x}) + f(\tau) \nu_{i}^{j}(\vec{x}).
\label{QI1}
\end{equation}
Inserting Eqs. (\ref{exp1}) and (\ref{exp2}) into Eq. (\ref{AN1})  and using the same parametrization 
of Eq. (\ref{PAR1A}) for the 
inhomogeneous contributions of the pressure and of the energy density the following pair of conditions can be obtained:
\begin{eqnarray}
&& \partial_{\tau}\biggl(\frac{\partial_{\tau} \beta}{2 a }\biggr) + \frac{{\mathcal H}}{a} \partial_{\tau} \beta = - a \ell_{\mathrm{P}}^2 \biggl( \frac{3 p^{(1)} + \rho^{(1)}}{3} 
+ \rho_{\mathrm{B}} \biggr),
\label{QI2}\\
&& \partial_{\tau} {\mathcal H} = - \frac{a^2 \ell_{\mathrm{P}}^2}{6} ( \rho^{(0)} + 3 p^{(0)}).
\label{QI3}
\end{eqnarray} 
Supposing, for sake of simplicity, that the barotropic index is constant we shall have that 
\begin{equation}
\rho^{(1)} = - \frac{1}{(3 w + 1)} \frac{[\partial_{\tau}^2 \beta + {\mathcal H} \partial_{\tau}\beta]}{\ell_{\mathrm{P}}^2 a^2} - \frac{2 \rho_{\mathrm{B}} }{( 3 w + 1)},
\label{QI4}
\end{equation}
where the contribution of the electric fields has been consistently neglected.
Inserting now Eqs. (\ref{exp1}) and (\ref{exp2}) into Eq. (\ref{AN3})
the following equations can be readily obtained:
\begin{eqnarray}
&& \partial_{\tau} \biggl( \frac{\partial_{\tau} \beta_{i}^{j}}{2 a} \biggr) + 
{\mathcal H} \frac{\partial_{\tau} \beta}{2 a} \delta_{i}^{j} + 
\frac{3 {\mathcal H}}{2 a } \partial_{\tau} \beta_{i}^{j} + a r_{i}^{j} = - a \ell_{\mathrm{P}}^2 \biggl[ \frac{p^{(1)} - \rho^{(1)}}{2} \delta_{i}^{j} + 
\Pi_{i}^{j}(B) - p_{\mathrm{B}} \delta_{i}^{j}\biggr],
\label{QI5}\\
&& \partial_{\tau} {\mathcal H} + 2 {\mathcal H}^2 = - 
\frac{\ell_{\mathrm{P}}^2 a^2}{2}  (p^{(0)} - \rho^{(0)}).
\label{QI6}
\end{eqnarray}
Note that, from Eqs. (\ref{QI3}) and (\ref{QI6}), the standard Hubble parameter is given by $H = H_{\mathrm{i}} a^{- 3(w +1)/2}$ where $H_{\mathrm{i}}$ is an integration constant with the same meaning of the analog constant introduced in section \ref{sec5}. Using then Eq. (\ref{QI4}) to eliminate $\rho^{(1)}$ from Eq. (\ref{QI5}), we have that
\begin{eqnarray}
&& \partial_{\tau}^2 \beta_{i}^{j} + 2 {\mathcal H} \partial_{\tau} \beta_{i}^{j} + 
\delta_{i}^{j} \biggl( \frac{1 - w}{1 + 3 w} \partial_{\tau}^2 \beta + 2 \frac{1+w}{1 + 3 w}
{\mathcal H} \partial_{\tau} \beta\biggr)  
\nonumber\\
&& = \frac{4}{3} a^2 \ell_{\mathrm{P}}^2 \biggl( \frac{ 3 w-1}{3 w+ 1}\biggr) \rho_{\mathrm{B}}  - 2 a^2 \ell_{\mathrm{P}}^2 - 2 a^2 r_{i}^{j}
\Pi_{i}^{j}(B).
\label{QI7}
\end{eqnarray}
Equation (\ref{QI7}) can be solved by positing
\begin{equation}
\beta_{i}^{j}(\vec{x},\tau) = g(\tau) \mu_{i}^{j}(\vec{x})+ f(\tau) \nu_{i}^{j}(\vec{x}).
\label{QI8}
\end{equation}
Then, inserting Eq. (\ref{QI8}) into Eq. (\ref{QI7}) and assuming $w \neq 1/3$ 
we obtain 
\begin{eqnarray}
&& g(\tau) = a^{3 w +1}, \qquad f(\tau) = a^{3 w -1}
\label{QI9}\\
&& \mu_{i}^{j}(\vec{x}) = - \frac{4}{H_{\mathrm{i}}^2 ( 3 w + 5) ( 3 w +1)} \biggl[ P_{i}^{j}(\vec{x}) 
+ \frac{3 w^2 - 6 w - 5}{4 ( 9 w + 5)} P(\vec{x}) \delta_{i}^{j} \biggr],
\label{QI10}\\
&& \nu_{i}^{j}(\vec{x}) = - \frac{\ell_{\mathrm{P}}^2}{\pi H_{\mathrm{i}}^2 ( 3 w -1) (3 w + 1)} \biggl[ B_{i}(\vec{x})B^{j}(\vec{x}) + \frac{9 w^2 - 9 w - 8}{9 ( 6 w + 5 - 3 w^2)} B^2(\vec{x}) \delta_{i}^{j} \biggr],
\label{QI11}
\end{eqnarray}
where $P_{i}^{j}(\vec{x}) = r_{i}^{j}(\vec{x},\tau) a^2(\tau)$ accounts for the 
intrinsic curvature computed from $\alpha_{ij}(\vec{x})$ and, by definition, 
\begin{equation}
B_{i}(\vec{x}) B^{j}(\vec{x}) = \frac{\alpha_{m i}(\vec{x})}{\alpha(\vec{x})} B^{m}(x) B^{j}(\vec{x}), \qquad 
B^2(\vec{x}) = \frac{\alpha_{m n}(\vec{x})}{\alpha(\vec{x})} B^{m}(\vec{x}) B^{n}(\vec{x}), 
\label{QI11a}
\end{equation}
with $\alpha(\vec{x}) = \mathrm{det}(\alpha_{ij})$.
Consequently $\gamma_{ij}(\vec{x},\tau)$ can be written as 
\begin{equation}
\gamma_{i j}(\vec{x},\tau) = a^2(\tau)\biggl[ \alpha_{ij}(\vec{x}) +  \mu_{ij}(\vec{x}) a^{3w+1} + \nu_{ij}(\vec{x}) a^{3 w-1}\biggr].
\label{QI12}
\end{equation}
If $w= 1/3$ 
\begin{equation}
\gamma_{ij}(\vec{x},\tau) = \tau^2 \biggl[\alpha_{ij}(\vec{x}) + \mu_{i j}(\vec{x}) \tau^2 + 
 \nu_{ij}(\vec{x}) \ln{\tau}\biggr],
\label{QI13}
\end{equation}
where, in the latter case, 
\begin{equation}
\mu_{i}^{j}(\vec{x})= - \frac{1}{3 H_{\mathrm{i}}^2} \biggl[ P_{i}^{j}(\vec{x}) - \frac{5}{24} P(\vec{x}) \delta_{i}^{j} \biggr],
\qquad \nu_{i}^{j}(\vec{x})  = - \frac{\ell_{\mathrm{P}}^2}{2 \pi H_{\mathrm{i}}^2} \biggl[ B_{i}(\vec{x}) B^{j}(\vec{x}) - 
\frac{B^2(\vec{x})}{3} \delta_{i}^{j} \biggr].
\label{QI14}
\end{equation}
If $w < -1/3$ the contributions of the spatial curvature and of the magnetic field become progressively subleading even if the terms proportional to $\nu_{ij}$ 
are diluted faster. If $ -1/3 < w< 1/3$ the contribution of the spatial curvature is progressively 
increasing while the contribution associated with the magnetic fields decreases. Finally if $w > 1/3$ both terms increase but at a different rate.
The solutions illustrated in the present section generalize the perturbative treatment where 
the large-scale magnetic fields are taken as a supplementary component in the linearized equations 
of the relativistic metric perturbations.  The linearized evolution of the curvature perturbations is usually described not in terms of $\gamma_{ij}$ but rather in terms of an appropriate gauge-invariant combination. 
It is therefore necessary to consider the interplay between the two 
different descriptions in the case when fully inhomogeneous magnetic fields contribute to the 
curvature perturbations which also receive an independent contribution from the standard adiabatic mode. 

\renewcommand{\theequation}{7.\arabic{equation}}
\setcounter{equation}{0}
\section{Nonlinear adiabatic mode in the plasma frame}
\label{sec7}
The initial conditions or the calculation of the CMB observables in the presence 
of large-scale magnetic fields are set within linear perturbation theory 
expressed either in a specific gauge or within a suitable 
gauge-invariant treatment \cite{mg1,mg2}.  In contrast with earlier 
studies, it has been recently clarified that different kinds of initial conditions 
are contemplated ranging from the magnetized adiabatic mode 
to the various magnetized entropy modes. Nonlinear generalizations of the 
magnetized curvature perturbations will now be discussed.

In the standard perturbative treatment of large-scale inhomogeneities,
the curvature perturbations on comoving 
orthogonal hypersurfaces (conventionally denoted by ${\mathcal R}$) and 
the curvature perturbations on uniform density hypersurfaces (conventionally 
denoted by $\zeta$) are often used to parametrize the large-scale 
curvature inhomogeneities. Both variables are gauge-invariant and can therefore 
be expressed in any gauge such as the longitudinal gauge 
 or the synchronous gauge \cite{bardeen,BERT}. 
In the longitudinal gauge (see appendix \ref{APPB} for an explicit definition in terms 
of the ADM variables) and in the plasma frame ${\mathcal R}$ and $\zeta$ can be written as\footnote{The notation $\overline{{\mathcal R}}$ and 
$\overline{\zeta}$ (as opposed to ${\mathcal R}$ and $\zeta$) is meant to distinguish the quantities 
defined in perturbation theory from their counterpart defined to a given order in the gradient expansion.}:
\begin{eqnarray}
\overline{{\mathcal R}}(\vec{x},\tau) &=& - \psi - \frac{{\mathcal H} ( {\mathcal H} \phi + \partial_{\tau} \psi)}{{\mathcal H}^2 - \partial_{\tau} {\mathcal H}},
\label{NL1}\\
\overline{\zeta}(\vec{x},\tau) &=& - \psi + \frac{\delta^{(\mathrm{L})}_{\mathrm{s}} \rho + \delta_{\mathrm{s}} \rho_{\mathrm{B}} }{3 (\rho_{\mathrm{t}} + 
p_{\mathrm{t}})}, 
\label{NL2}
\end{eqnarray}
where $\delta^{(\mathrm{L})}_{\mathrm{s}} \rho$ denotes the scalar fluctuation of the energy density 
of the fluid sources in the longitudinal gauge; $\delta_{\mathrm{s}} \rho_{\mathrm{B}} = B^2(\vec{x})/[8 \pi a^4(\tau)]$ 
denotes the fluctuation of the energy density associated with the fully inhomogeneous magnetic field, 
while $\rho_{\mathrm{t}}$ and $p_{\mathrm{t}}$ are, respectively, the total energy density and pressure 
determining the (homogeneous and isotropic) background geometry. 

By computing the difference $(\overline{\zeta} - \overline{{\mathcal R}})$ and by comparing the obtained 
result with the Hamiltonian constraint  (see Eq. (\ref{bard5}) of appendix \ref{APPB}), 
the difference $(\overline{\zeta} - \overline{{\mathcal R}})$ becomes
\begin{equation}
\overline{\zeta} - \overline{{\mathcal R}} = \frac{2 \nabla^2 \psi}{ 3\,\ell_{\mathrm{P}}^2\, a^2 (p_{\mathrm{t}} + \rho_{\mathrm{t}})}.
\label{NL4}
\end{equation}
which is negligible in the limit 
of typical length-scales much larger than the Hubble radius at the 
corresponding epoch. The variables $\overline{\zeta}$ and $\overline{{\mathcal R}}$ are gauge-invariant: their numerical 
value does not change for two coordinate systems connected by infinitesimal coordinate 
transformations of the type 
\begin{equation}
x^{\mu} \to \tilde{x}^{\mu} = x^{\mu} + \epsilon^{\mu}, \qquad \epsilon_{\mu} = a^2(\tau)(\epsilon_{0},- \partial_{i} \epsilon)
\label{NL4a}
\end{equation}
where both $\epsilon_{0}$ and $\epsilon$ depend on the space-time point\footnote{
As explained in appendix \ref{APPB}, the first part of the present section assumes the standard perturbative treatment and, therefore, the underlying background geometry is taken to be conformally flat as in the case of the vanilla $\Lambda$CDM scenario.}. A coordinate transformation of the type (\ref{NL4a}) will 
change metric fluctuations according to the Lie derivative in the direction of the vector $\epsilon^{\mu}$; 
consider, for sake of concreteness, the transformation from the synchronous coordinate 
system to the longitudinal one:
\begin{eqnarray}
&& \psi^{(\mathrm{S})} \to \psi^{(\mathrm{L})} = \psi^{(\mathrm{S})} + {\mathcal H} \epsilon_{0}, \qquad 
 F^{(\mathrm{S})} \to F^{(\mathrm{L})} = F^{(\mathrm{S})} - \epsilon,
\label{GT1}\\
&& B^{(\mathrm{S})} \to B^{(\mathrm{L})} = B^{(\mathrm{S})} + \epsilon_{0} - \partial_{\tau} \epsilon,\qquad 
\phi^{(\mathrm{S})} \to \phi^{(\mathrm{L})} = \phi^{(\mathrm{S})} - \partial_{\tau} \epsilon_{0} - {\mathcal H} \epsilon_{0}.
\label{GT2}
\end{eqnarray}
In the synchronous gauge $\phi^{(\mathrm{S})} =0$ and $B^{(\mathrm{S})} =0$ while, 
by definition of longitudinal gauge, $F^{(\mathrm{L})} =0$ and $B^{(\mathrm{L})} =0$ (see, e.g. \cite{BERT}). Consequently, the standard relation between the variables appearing in $\overline{\zeta}$ and $\overline{{\mathcal R}}$ can be written as 
\begin{equation}
\psi = - \xi + {\mathcal H} \partial_{\tau} F,\qquad \phi = - \partial_{\tau}^2 F - {\mathcal H} \partial_{\tau} F,
\qquad \delta_{\mathrm{s}}^{(\mathrm{L})}= \delta_{\mathrm{s}}^{(\mathrm{S})} - \partial_{\tau} \rho_{\mathrm{t}} \partial_{\tau} F.
\label{GT5}
\end{equation}
By inserting Eq. (\ref{GT5}) into Eqs. (\ref{NL1}) and (\ref{NL2}), the standard synchronous 
 expression of  $\overline{{\mathcal R}}$ and $\overline{\zeta}$ can be readily obtained 
\begin{equation}
\overline{{\mathcal R}} = \xi + \frac{{\mathcal H} \partial_{\tau} \xi}{{\mathcal H}^2 - \partial_{\tau} {\mathcal H}}, \qquad 
\overline{\zeta}  = \xi + \frac{\delta^{(\mathrm{S})}_{\mathrm{s}} \rho + \delta_{\mathrm{s}} \rho_{\mathrm{B}} }{3 (\rho_{\mathrm{t}} + 
p_{\mathrm{t}})},
\label{GT5a}
\end{equation}
and it coincides, as expected, with previous discussions (see, e.g. \cite{BERT}). Because 
of gauge invariance, the evolution equations for $\overline{\zeta}$ and $\overline{{\mathcal R}}$ can be 
derived and discussed in any gauge, for instance, by linearizing the covariant conservation 
of the energy momentum tensor of the sources. Let us consider, for sake of concreteness, the 
matter radiation transition and let us define the fluctuation of the total pressure in terms 
of the sound speed and of the non-adiabatic pressure fluctuation $\delta p_{\mathrm{nad}}$:
\begin{equation}
\delta_{\mathrm{s}} p = \overline{c}_{\mathrm{st}}^2 \delta_{\mathrm{s}} \rho + \delta p_{\mathrm{nad}}, \qquad \overline{c}_{\mathrm{st}}^2 = \frac{\partial_{\tau} p_{\mathrm{t}}}{\partial_{\tau} \rho_{\mathrm{t}}}  = 
\frac{4}{3 ( 3 \alpha + 4)}
\label{zetaA}
\end{equation}
where $\alpha = a(\tau)/a_{\mathrm{eq}}$ and the sound speed, as shown in the case of the matter radiation 
transition, is fully homogeneous.  The result for the evolution of $\overline{\zeta}$ when 
the plasma effects are carefully included has been given in the second paper of Ref. \cite{MGL}
\begin{eqnarray}
\partial_{\tau}\overline{\zeta} + \frac{{\mathcal H} \delta p_{\mathrm{nad}}}{(p_{\mathrm{t}} + \rho_{\mathrm{t}})} = \frac{\vec{E} \cdot \vec{J}}{3 a^4 (p_{\mathrm{t}} + \rho_{\mathrm{t}})} + \frac{{\mathcal H}\delta_{\mathrm{s}} \rho_{\mathrm{B}} (3 \overline{c}_{\mathrm{st}}^2 -1)}{ 3 (\rho_{\mathrm{t}} + p_{\mathrm{t}})} 
+  \frac{{\mathcal H} \delta_{\mathrm{s}} \rho_{\mathrm{E}}}{ 3 (\rho_{\mathrm{t}} + p_{\mathrm{t}})} {\mathcal F}(\alpha, \overline{c}_{\mathrm{st}}) - \frac{\theta_{\mathrm{t}}}{3},
\label{zeta1}
\end{eqnarray}
where the following quantities have been introduced
\begin{equation}
{\mathcal F}(\alpha, \overline{c}_{\mathrm{st}}) =  \frac{[3 \overline{c}_{\mathrm{st}}^2 g_{1}^2 \alpha^2 + g_{2}^2 ( 3\overline{c}_{\mathrm{st}}^2 -2)]}{g_{1}^2 \alpha^2 +g_{2}^2}, \qquad 
\theta_{\mathrm{t}}=\vec{\nabla} \cdot \vec{v}_{\mathrm{t}};
\label{zeta2}
\end{equation}
$g_{1}$ and $g_{2}$ are  two numerical constants which depend upon the $\Lambda$CDM parameters 
which can be explicitly computed (see second paper of \cite{MGL}). For the fiducial set of 
$\Lambda$CDM parameters $g_{1} \ll g_{2} \simeq 10^{-5}$.
Neglecting the electric fields, Eq. (\ref{zeta2}) 
can be easily solved by direct integration since, for length-scales much larger than the Hubble radius, 
Eq. (\ref{zeta2}) reads:
\begin{equation}
 \frac{\partial \overline{\zeta}}{\partial \ln{\alpha}} = -
\frac{\delta_{\mathrm{s}} p_{\mathrm{nad}}}{(p_{\mathrm{t}} + \rho_{\mathrm{t}})} + \biggl(\overline{c}_{\mathrm{st}}^2 - \frac{1}{3}\biggr) \frac{\delta_{\mathrm{s}} \rho_{\mathrm{B}}}{(p_{\mathrm{t}} + \rho_{\mathrm{t}})},
\label{zeta3}
\end{equation}
with the result that, in the case of the magnetized adiabatic mode (i.e.  $\delta p_{\mathrm{nad}} =0$), the solution 
of Eq. (\ref{zeta3}) becomes: 
\begin{equation}
\overline{\zeta}(\vec{x},\tau) = \overline{\zeta}_{\ast}(\vec{x}) - \frac{3}{4} R_{\gamma} \Omega_{\mathrm{B}}(\vec{x}) \frac{\alpha}{3 \alpha + 4}.
\label{zeta4}
\end{equation}
In the nonlinear case it does not make sense 
to separate the energy density and the pressure in a background value supplemented 
by the corresponding fluctuations. Still it will be possible to define 
the sound speed in terms of the barotropic index, i.e.
\begin{equation}
c_{\mathrm{st}}^2(\vec{x},\tau)= \frac{\partial_{\tau} p}{\partial_{\tau} \rho} = w + \frac{\partial_{\tau} w}{N K ( p + \rho)}.
\label{zeta6}
\end{equation}
Equation  (\ref{zeta6})  coincides with the second relation of Eq. (\ref{zetaA}) in the fully homogeneous case. 
However,  when the description of the geometry is given in fully inhomogeneous terms, 
the two definitions lead to different results. 

Let us now see how the curvature perturbations can be generalized to nonlinear level.
A possible nonlinear generalization of the the variable $\zeta$  is given by:
\begin{equation}
\zeta_{i} = \frac{\nabla_{i} \ln{\sqrt{\gamma}}}{3} + \frac{N K}{3 \partial_{\tau} \rho} 
\nabla_{i}( \rho + \rho_{\mathrm{B}}),
\label{NL6a}
\end{equation}
while a complementary generalization is 
\begin{equation}
\zeta_{i} = \frac{\nabla_{i} \ln{\sqrt{\gamma}}}{3} + \frac{\nabla_{i}( \rho + \rho_{\mathrm{B}}) }{3 ( p + \rho)}. 
\label{NL6}
\end{equation}
Note that Eqs. (\ref{NL6a}) and (\ref{NL6}) are equivalent to zeroth order in the gradient expansion since 
$\partial_{\tau} \rho = N K (p + \rho)$. However, for practical reasons, the definition (\ref{NL6}) 
will be preferred.  With similar logic, the nonlinear generalization of ${\mathcal R}$ can be written as 
\begin{equation}
{\mathcal R}_{i} = \frac{ \nabla_{i} \ln{\sqrt{\gamma}}}{3} + \frac{\nabla_{i} ( K^2 - \mathrm{Tr} K^2)}{6 \ell_{\mathrm{P}}^2 ( \rho + p )}. 
\label{NL7}
\end{equation}
The variable given in Eq. (\ref{NL6a}) is inspired by the one defined in Ref. \cite{shell}
in the case of an energy-momentum tensor dominated by a single scalar field. 
The pair of variables defined in Eqs. (\ref{NL6}) and (\ref{NL7}) can be compared to their 
linearized counterpart by expressing the extrinsic curvature in a specific gauge. For instance, in the longitudinal gauge, using the results of appendix \ref{APPB} (and, in particular, Eq. (\ref{bard2})) Eqs. (\ref{NL6}) and (\ref{NL7}) become
\begin{equation}
\zeta_{i} \to \nabla_{i}\biggl[ - \psi + \frac{\delta^{(\mathrm{L})}_{\mathrm{s}} \rho + \delta_{\mathrm{s}} \rho_{\mathrm{B}}}{3 (\rho_{\mathrm{t}} + 
p_{\mathrm{t}})}\biggr],\qquad 
{\mathcal R}_{i} \to \nabla_{i} \biggl[- \psi - \frac{{\mathcal H} ( {\mathcal H} \phi + \partial_{\tau} \psi)}{{\mathcal H}^2 - \partial_{\tau} {\mathcal H}}\biggr],
\label{NL7A}
\end{equation}
which also implies, because of Eqs. (\ref{NL1}) and (\ref{NL2}), $\zeta_{i}^{(\mathrm{L})} \to \nabla_{i} \overline{\zeta}$ and  ${\mathcal R}_{i}^{(\mathrm{L})} \to \nabla_{i} \overline{{\mathcal R}}$. By taking the difference of $\zeta_{i}$ and ${\mathcal R}_{i}$ and by using the Hamiltonian constraint of Eq. (\ref{EE6}) we obtain
\begin{equation}
\zeta_{i} - {\mathcal R}_{i} =  \frac{\nabla_{i} r}{6 \ell_{\mathrm{P}}^2 ( p + \rho)} - \frac{\nabla_{i} ( p + \rho) u^2}{3 ( p + \rho)}.
\label{NL10}
\end{equation}
By expanding the right hand side of Eq. (\ref{NL10}) to first order in the (longitudinal) 
metric fluctuations, it can be verified that the obtained result coincides with the spatial gradient of Eq. 
(\ref{NL4}).  As far as gauge-invariance is concerned, the approach followed here is  similar to the one invoked 
in related contexts (see, e.g. \cite{shell,salopek}):  Eqs. (\ref{NL6}) and (\ref{NL7}) can be shown to be gauge invariant to a given order in the gradient expansion. Suppose, indeed, to evaluate Eqs. (\ref{NL6}) and (\ref{NL7}) 
not in the longitudinal gauge but in a different gauge, for instance 
the synchronous gauge. If the variables are truly nonlinear gauge-invariant variables 
they must also be gauge-invariant for infinitesimal gauge transformations of the kind discussed 
in Eqs. (\ref{GT1})--(\ref{GT2}).  With the help 
of Eqs. (\ref{bard9}), the synchronous gauge expression of ${\mathcal R}_{i}$ can be written as:
\begin{equation}
{\mathcal R}^{(\mathrm{S})}_{i} = \nabla_{i} \biggl[ \xi + \frac{{\mathcal H} \partial_{\tau} \xi}{{\mathcal H}^2 - \partial_{\tau} {\mathcal H}}\biggr] + \frac{1}{3} \nabla_{i} \biggl[ \nabla^2 F + \frac{{\mathcal H}}{{\mathcal H}^2 - \partial_{\tau} {\mathcal H}} \nabla^2 \partial_{\tau}  F \biggr],
\label{NL11}
\end{equation}
where the notation ${\mathcal R}^{(\mathrm{S})}_{i}$ reminds that ${\mathcal R}_{i}$ 
is computed in the synchronous coordinate system. The first term in squared brackets of Eq. (\ref{NL11}) 
clearly coincides with the spatial gradient of $\overline{{\mathcal R}}$ (see Eq. (\ref{NL1})).  
The second term in squared bracket appearing in Eq. (\ref{NL11}) breaks therefore gauge-invariance 
but it is of higher order in the gradients. The same kind of considerations can also be extended to the case of $\zeta_{i}$. In summary, recalling Eq. (\ref{GT5a}) (i.e. the synchronous form of $\overline{{\mathcal R}}$ and $\overline{\zeta}$), 
we have that, in the synchronous parametrization,  
\begin{eqnarray}
\zeta_{i}^{(\mathrm{S})} &=& \nabla_{i} \overline{\zeta} + \frac{\nabla_{i} \nabla^2 F}{3}  \equiv \zeta_{i}^{(\mathrm{L})} + {\mathcal O}(\nabla^3),
\label{NL12}\\
{\mathcal R}_{i}^{(\mathrm{S})} &=& \nabla_{i} \overline{{\mathcal R}} + \frac{1}{3} \nabla_{i} \biggl[ \nabla^2 F + \frac{{\mathcal H}}{{\mathcal H}^2 - \partial_{\tau} {\mathcal H}} \nabla^2 \partial_{\tau}  F \biggr]\equiv 
{\mathcal R}_{i}^{(\mathrm{L})} + {\mathcal O}(\nabla^3).
\label{NL13}
\end{eqnarray}
Equations (\ref{NL12}) and (\ref{NL13}) are rather interesting and suggest that 
the variables $\zeta_{i}$ and ${\mathcal R}_{i}$ defined in Eqs. (\ref{NL6}) and (\ref{NL7}) 
can be computed in any suitable gauge. If the two gauges differ by an infinitesimal coordinate transformation, the 
terms leading to a breaking of gauge-invriance are of higher order in the gradient expansion.
In the specific example of Eqs. (\ref{NL12}) and (\ref{NL13}) the extra terms contain three 
gradients. 

The observation of the previous paragraph fails, however, in the case of more general coordinate transformations. 
Still the results suggested by Eqs. (\ref{NL12}) and (\ref{NL13}) can be generalized along the 
lines of \cite{shell,salopek} to the case when the coordinate transformations are finite 
but still sufficiently well behaved. Consider, in particular, the following 
coordinate transformation 
\begin{equation}
X^{i}= X^{i} ( \vec{x},\tau),\qquad T= T(\vec{x},\tau).
\label{NL14}
\end{equation}
As discussed in \cite{shell,salopek}, along lines of constant $X^{i}$ the old coordinates change 
as $d x^{\mu} = \partial^{\mu} T d s$ where $s$ is an arbitrary parameter and 
\begin{equation}
X^{i} = f^{i}(\vec{x}) + \int \frac{\partial^{i} T}{\partial_{\tau} T \partial^{\tau} T} dT .
\label{NL15}
\end{equation}
If the new time coordinate $T$ is non singular the second term in the transformation 
of Eq. (\ref{NL15}) can be discarded to leading order in the gradient expansion. 

From Eqs. (\ref{NL6}) and (\ref{NL7}) the evolution equation of the nonlinear 
curvature perturbations can be obtained in explicit terms by using the governing 
equations and the covariant to conservation equation to leading order in the 
spatial gradients. Let us therefore take the first time derivative of $\zeta_{i}$ and let us 
drop the contribution of the electric fields which are subleading to zeroth order 
in the drift approximation; the result of this simple manipulation is:
\begin{equation}
\partial_{\tau} \zeta_{i} = \frac{1}{3} \nabla_{i}\biggl(\frac{\partial_{\tau} \sqrt{\gamma}}{\sqrt{\gamma}}\biggr) + \frac{1}{3( p + \rho)} \nabla_{i} \bigl( \partial_{\tau} \rho + \partial_{\tau} \rho_{\mathrm{B}}\bigr)
- \frac{\nabla_{i} (\rho + \rho_{\mathrm{B}})}{3 ( p + \rho)^2}[ \partial_{\tau} p + \partial_{\tau} \rho].
\label{NL15a}
\end{equation}
But to first order in the gradient expansion $\partial_{\tau} \rho = N K (p + \rho)$ and 
$\partial_{\tau} \rho_{\mathrm{B}} = 4 N K \rho_{\mathrm{B}}/3$. Thus Eq. (\ref{NL15a}) becomes:
\begin{equation}
\partial_{\tau} \zeta_{i} = \frac{K N}{3 ( p + \rho)} \biggl(\nabla_{i} p - c_{\mathrm{st}}^2 \nabla_{i} \rho\biggr) 
+ \frac{N K}{3} \biggl( \frac{1}{3} - c_{\mathrm{st}}^2\biggr) \frac{\nabla_{i} \rho_{\mathrm{B}}}{(\rho + p)}
+ \frac{4}{9} \frac{\rho_{\mathrm{B}}}{p + \rho} \nabla_{i} ( N K).
\label{NL16}
\end{equation}
Equation (\ref{NL16}) generalizes Eq. (\ref{zeta1}) as it can be easily appreciated 
by considering various specific limits. Suppose, for instance, that the barotropic index $w$ is constant both in time and in space and that $\rho_{\mathrm{B}} =0$. Then, $c_{\mathrm{st}}^2 = w$ and 
Eq. (\ref{NL16}) implies $\partial_{\tau} \zeta_{i} =0$. This is the case of the 
single adiabatic mode in the absence of magnetic fields. 

In general terms the solution of Eq. (\ref{NL16}) represents a complicated self-consistent 
problem since both 
$c_{\mathrm{st}}^2$ and $w$ will be both space and time dependent; moreover, always 
from a general point of view, $\nabla_{i} ( N K)$ and $\nabla_{i} \rho_{\mathrm{B}}$ can well be of the same order. To simplify the situation let us make the following (not completely realistic) assumption 
\begin{equation}
\frac{\nabla_{i} (N K) }{NK} \ll \frac{\nabla_{i} \rho_{\mathrm{B}}}{\rho_{\mathrm{B}}}, \qquad 
w(\alpha) = \frac{1}{3 (\alpha + 1)},
\label{NL17}
\end{equation}
where, as before, $\alpha = a/a_{\mathrm{eq}}$.
Equation (\ref{NL16}) can then be solved directly and the result is:
\begin{equation}
\zeta_{i}(\vec{x}, \alpha) = \overline{\zeta}_{i}(\vec{x}) - \frac{3 \alpha}{4( 3 \alpha + 4)}\frac{\nabla_{i} \rho + \nabla_{i} \rho_{\mathrm{B}}}{\rho}, 
\label{NL18}
\end{equation}
where the integration constant has been matched with the value of $\zeta_{i}$ determined 
in the linearized approximation. The solution (\ref{NL18}) is interesting but it assumes 
an expression for the evolution of the barotropic index which is only justified 
in the homogeneous and isotropic case.

A much safer approach for the computation of the curvature perturbations is to first obtain the quasi-isotropic 
MHD solutions  at the wanted order in the gradients (as already done in section \ref{sec6}) and then to evaluate explicitly ${\mathcal R}_{i}$ and $\zeta_{i}$. 
Bearing in mind that, in the notations of section \ref{sec6}, 
$\sqrt{\gamma} = a^3(\tau) \sqrt{\alpha(\vec{x})} [ 1 + \beta(\vec{x},\tau)/2]$,
the explicit expression of ${\mathcal R}_{i}(\vec{x},\tau)$ can be computed and it is:
\begin{equation}
{\mathcal R}_{i}(\vec{x},\tau) = \frac{1}{6} \frac{\partial_{i} \alpha}{\alpha} + 
\frac{1}{6}  \biggl[ g(\tau) \partial_{i} \mu + f(\tau) \partial_{i} \nu\biggr]+ \frac{{\mathcal H}}{6 ({\mathcal H}^2 - \partial_{\tau} {\mathcal H})} \biggl[ \partial_{\tau} g \partial_{i} \mu 
+ \partial_{\tau} f \partial_{i} \nu \biggr].
\label{RC1}
\end{equation}
Recalling the explicit expressions of $\mu(\vec{x})$ and $\nu(\vec{x})$ we have that 
\begin{eqnarray}
{\mathcal R}_{i}(\vec{x},\tau) &=& \frac{1}{6} \frac{\partial_{i} \alpha}{\alpha}  - \frac{\partial_{i} P}{18 H_{\mathrm{i}}^2 (w + 1)}
a^{3 w +1} 
\nonumber\\
&-& \frac{ 4 ( 9 w + 7) ( 9 w + 1)}{ 3 ( 3 w -1) ( 3 w+1) ( 6 w + 5 - 3 w^2) ( w + 1)} \partial_{i} \Omega_{\mathrm{B}} a^{ 3 w -1},
\label{RC2}
\end{eqnarray} 
where now $\Omega_{\mathrm{B}}(\vec{x}) = \ell_{\mathrm{P}}^2\, B^2(\vec{x})/(24 \pi H_{\mathrm{i}}^2)$.
By using the explicit expression of the Hamiltonian constraint the variable $\zeta_{i}(\vec{x},\tau)$ 
is only sensitive, by construction, to the gradients of the magnetic fields, i.e. 
\begin{eqnarray}
\zeta_{i}(\vec{x},\tau) &=& {\mathcal R}_{i}(\vec{x},\tau) + \frac{\partial_{i} P}{6 \ell_{\mathrm{P}}^2 a^2 (\rho + p)}
\nonumber\\
&=& \frac{1}{6} \frac{\partial_{i} \alpha}{\alpha} - \frac{ 4 ( 9 w + 7) ( 9 w + 1)}{ 3 ( 3 w -1) ( 3 w+1) ( 6 w + 5 - 3 w^2) ( w + 1)} \,\,\partial_{i} \Omega_{\mathrm{B}} a^{ 3 w -1}.
\label{RC3}
\end{eqnarray}
Both in the case of Eq. (\ref{RC2}) as well as in the case of Eq. (\ref{RC3}) the leading order result is 
fully inhomogeneous  and it accounts for the large-scale curvature perturbations connected, in this specific 
case, to the adiabatic solution. The spatial gradients affecting
${\mathcal R}_{i}$ are induced both by the spatial curvature as well as by the magnetic inhomogeneities. 
Conversely, the spatial gradients of the curvature affect $\zeta_{i}$ only to higher order because of a cancellation 
arising when the solution for ${\mathcal R}_{i}$ is inserted into the fully ingomogeneous form 
of the Hamiltonian constraint of Eq. (\ref{NL10}).
Note that in the case $w=1/3$, Eq. (\ref{QI14}) implies that $\nu_{i}^{j}$ is traceless. This implies that
the contribution of the magnetic fields to the curvature perturbation vanishes to first-order in the 
gradient expansion in a radiation-dominated Universe while the effect of ordinary spatial 
inhomogeneities does contribute.  Consequently, it has been shown that while it is difficult to solve explicitly 
the nonlinear generalization of the evolution equation for the magnetized curvature 
perturbations, the same techniques leading to the 
quasi-isotropic solutions can be used to derive the expression of the nonlinear 
generalization of the magnetized curvature perturbations.

\renewcommand{\theequation}{8.\arabic{equation}}
\setcounter{equation}{0}
\section{Concluding remarks}
\label{sec8}
The general relativistic gradient expansion has been combined 
with the standard tenets of the drift approximation used in the description of cold plasmas.
Nonlinear effects are typical both of general relativistic dynamics as well as of magnetohydrodynamics 
in flat space-time. It is then natural to treat them in a unified perspective where the standard 
linearized approximation is not assumed from the very beginning.
A fully nonlinear system has been derived in terms of the ADM variables and discussed in specific 
physical limits. The evolution of magnetic and curvature inhomogeneities 
has been treated and solved on the same footing. The present findings have also been contrasted with the standard linearized description of magnetized curvature perturbations both in the presence of non-adiabatic pressure fluctuations as well as in the case of adiabatic initial conditions. After introducing the nonlinear magnetized adiabatic mode, an explicit 
evolution equation for the magnetized curvature perturbations has been derived to leading order in the spatial 
gradients. While the results of the present investigation pave the way for a more thorough scrutiny of nonlinear effects in gravitating  plasmas prior to photon decoupling, they also fill an existing gap in the present literature.
Indeed, the treatment of magnetized plasmas usually rests upon equations written in homogeneous and isotropic backgrounds supplemented by the corresponding relativistic fluctuations of the geometry and of the fluid sources. Here the modest attempt has been to suggest and partially develop a description which is independent of the background but valid to a given order in the spatial gradients.
\newpage
\begin{appendix}
\renewcommand{\theequation}{A.\arabic{equation}}
\setcounter{equation}{0}
\section{Electromagnetic fields and ADM decomposition}
\label{APPA}
The explicit form of the of the Christoffel symbols in terms of the ADM variables of Eq. (\ref{ADM1}) are:
\begin{eqnarray}
\Gamma_{0 \, 0}^{0}&=& \frac{\partial_{\tau}N}{N} + \frac{N^{k} \, ^{(3)}\nabla_{k} N}{N} -\frac{N^{m} N^{n}}{N} K_{m n},
\nonumber\\
\Gamma_{0\, i}^{0} &=& \frac{\nabla_{i} N}{N} - \frac{N^{m}}{N} K_{m i},
\nonumber\\
\Gamma_{00}^{i} &=&  N\, ^{(3)}\nabla^{i} N + \partial_{\tau}N^{i} - 2 N N^{m} K_{m}^{i} + N^{m}\,  ^{(3)}\nabla_{m} N^{i}
\nonumber\\
&-& \frac{N^{i}}{N}\biggl( \partial_{\tau} N +
N^{m} \,^{(3)}\nabla_{m} N - N^{m} N^{n} K_{m n}\biggr),
\nonumber\\
\Gamma_{ij}^{0} &=& - \frac{1}{N} K_{i j},
\nonumber\\
\Gamma_{i 0}^{j} &=& - N^{j} \frac{^{(3)}\nabla_{i}N}{N} + \frac{N^{j} N^{m}}{N} K_{m i} - N  K_{i}^{j} + ^{(3)}\nabla_{i} N^{j},
\nonumber\\
 \Gamma_{m n}^{i} &=& ^{(3)}\Gamma_{m n}^{i} + \frac{N^{i}}{N} K_{m n}.
 \label{ADM1c}
\end{eqnarray}
In Eq. (\ref{ADM1c}) the indices are lowered and raised ny using $\gamma_{ij}$ so that, for instance, 
$K_{i}^{j} = \gamma^{k j} K_{k i}$; $^{(3)}\nabla_{i}$ is the covariant derivative defined with respect to the spatial metric $\gamma_{ij}$ and the corresponding Christoffel symbol is: 
\begin{equation}
^{(3)}\Gamma_{m n}^{i}= \frac{1}{2} \gamma^{i j}\biggl( - \partial_{j} \gamma_{m n} + 
\partial_{n} \gamma_{j m} + \partial_{m} \gamma_{n j} \biggr).
\label{ADM1ca}
\end{equation}
Recalling Eqs. (\ref{ADM1a}) and (\ref{ADM1b})  the traces $K$, $\mathrm{Tr}K^2$ and $r$ are defined as:
\begin{equation}
K = K_{i}^{i},\qquad \mathrm{Tr}K^2= K_{i}^{j} \, K_{j}^{i},\qquad r = \gamma^{ij} r_{ij}.
\label{ADM6}
\end{equation}
From Eq. (\ref{ADM1c}) the components of the Ricci tensor and the Ricci scalar read
\begin{eqnarray}
R_{0}^{0} &=&  
\frac{\partial_{\tau} K}{N}- \mathrm{Tr}K^2 + \frac{\nabla^2 N}{N} - \frac{N^{m}}{N} \nabla_{m} K + \frac{N^{q}}{N} {\mathcal L}_{q},
\label{ADM7}\\
R_{i}^{0} &=& \frac{1}{N} {\mathcal L}_{i},
\label{ADM8}\\
R_{i}^{j} &=& \frac{1}{N} \partial_{\tau} K_{i}^{j} - K K_{i}^{j} - r_{i}^{j} + \frac{1}{N} \nabla_{i} \nabla^{j} N - \frac {N^{m}}{N} \nabla_{m} K_{i}^{j}
\nonumber\\
&+& \frac{1}{N} \nabla_{m} N^{j} K^{m}_{i}-  \frac{1}{N} \nabla_{i} N^{m} K_{m}^{j} - \frac{N^{j}}{N} {\mathcal L}_{i},
\label{ADM9}
\end{eqnarray}
where, for sake of conciseness, the following notations have been adopted:
\begin{equation}
^{(3)} \nabla_{i} = \nabla_{i}, \qquad {\mathcal L}_{i} = \biggl(\nabla_{i} K - \nabla_{k} K^{k}_{i} \biggr).
\label{ADM10} 
\end{equation}
From Eqs. (\ref{ADM7}), (\ref{ADM8}), (\ref{ADM9}) and (\ref{ADM10}) the components of the Einstein tensors can be easily obtained and they are:
\begin{eqnarray}
G_{0}^{0} &=& \frac{1}{2}\biggl(K^2 + r - \mathrm{Tr}K^2\biggr) + \frac{N^{q}}{N} {\mathcal L}_{q},
\label{ADM11}\\
G_{i}^{0} &=& \frac{1}{N} \biggl( \nabla_{i} K - \nabla_{k} K^{k}_{i} \biggr), 
\label{ADM12}\\
G_{i}^{j} &=& \frac{1}{N} \partial_{\tau} \biggl(K_{i}^{j} - K\delta_{i}^{j} \biggr) - K \biggl(K_{i}^{j} - \frac{K}{2} \delta_{i}^{j}\biggr)
+ \frac{\mathrm{Tr}K^2}{2} \delta_{i}^{j} 
\nonumber\\
&-& \biggl( r_{i}^{j} - \frac{r}{2} \delta_{i}^{j}\biggr) + \frac{1}{N} \biggl( \nabla_{i} \nabla^{j} N - \nabla^2 N \delta_{i}^{j} \biggr) 
- \frac{N^{m}}{N} \nabla_{m} \biggl( K_{i}^{j} - K \delta_{i}^{j} \biggr) 
\nonumber\\
&+&  \frac{1}{N} \nabla_{m} N^{j} K^{m}_{i} - 
\frac{1}{N} \nabla_{i} N^{m} K_{m}^{j} - \frac{N^{j}}{N} {\mathcal L}_{i}. 
\label{ADM13}
\end{eqnarray}
For a perfect relativistic fluid the components of the energy-moementum tensor with covariant indices are:
\begin{eqnarray}
&& T_{00} = (p + \rho) u_{0} u_{0} - p (N^2 - N_{k} N^{k}), 
\nonumber\\
&& T_{i0} = (p+ \rho) u_{0} u_{i} + p N_{i}, 
\nonumber\\ 
&& T_{ij} = (p + \rho) u_{i} u_{j} + p \gamma_{ij}.
\label{T1}
\end{eqnarray}
The condition $g^{\mu\nu} u_{\mu} u_{\nu} =1$ implies, in terms of the ADM variables,  
\begin{equation}
\biggl[ u_{0} - N^{k} u_{k}\biggr]^2 = N^2 ( 1 + u^2),\qquad u^2 = \gamma^{ij} u_{i} u_{j}.
\label{T2}
\end{equation}
Equations Eqs. (\ref{T1}) and (\ref{T2}) are written in general terms and, therefore,  we shall also have that, in general:
\begin{eqnarray}
&& T_{0}^{0} = \rho + ( p + \rho) \, u\, \sqrt{1 + u^2} \biggl[ \frac{N^{k} u_{k}}{N u}  + \frac{u}{\sqrt{ 1 + u^2}}\biggr],
\label{T3}\\
&& T_{i}^{0} = \frac{(p + \rho)}{N} \, \sqrt{ 1 + u^2} u_{i},
\label{T4}\\
&& T_{i}^{j} = - p \delta_{i}^{j} - (p + \rho) u_{i} u^{j} - \frac{p + \rho}{N} \sqrt{1 + u^2}\, u_{i}\, N^{j}.
\label{T5}
\end{eqnarray}
From Eqs. (\ref{ADM12}) and (\ref{T4}) the explicit form of the the Hamiltonian and of the momentum
constraints, (i.e. $G_{0}^{0} = \ell_{\mathrm{P}}^2 T_{0}^{0}$ and $G_{i}^{0} = \ell_{\mathrm{P}}^2 T_{i}^{0}$) becomes, 
respectively:
\begin{equation}
K^2 + r - \mathrm{Tr} K^2 = 2 \ell_{\mathrm{P}}^2 [ \rho + (\rho + p) u^2], \qquad {\mathcal L}_{i} = \ell_{\mathrm{P}}^2 
( p + \rho) \sqrt{1 + u^2} u_{i}.
\label{T5a}
\end{equation}
Note that the term proportional to ${\mathcal L}_{q}$ in $G_{0}^{0}$ vanishes exactly with the terms containing $N^{k}u_{k}$ 
in Eq. (\ref{T3}) once the momentum constraint is imposed.  
In the case $N_{i}= 0$ and within the parametrization of Eqs. (\ref{EL6a}) and (\ref{EL7a}) the relevant 
component of the fluid energy-momentum tensor can be written as
\begin{eqnarray}
T^{00} &=& \frac{1}{N^2} [ \rho \cosh^2{y} + p \sinh^2{y}],\qquad T^{0 i} = \frac{ p + \rho}{N} \sinh{y} \cosh{y} \hat{v}^{i},
\nonumber\\
T^{ij} &=& ( p+ \rho) \hat{v}^{i} \hat{v}^{j} \sinh^2{y} + p \gamma^{ij}.
\label{T8}
\end{eqnarray}
The covariant conservation of the energy-momentum tensor can be written in terms 
of the Christoffel symbols obtained in Eq. (\ref{ADM1c}) 
\begin{eqnarray}
\nabla_{\mu} T^{\mu 0}  &=& \partial_{0} T^{00} + \partial_{k} T^{k 0}+ \biggl( 2 \Gamma_{0 0}^{0} + \Gamma_{0 k}^{k} \biggr)\, T^{00} + 
\biggl( 3 \Gamma^{0}_{0 k} + \Gamma^{j}_{k j} \biggr)  T^{0 k} + \Gamma^{0}_{k j} T^{k j}, 
\label{T6}\\
\nabla_{\mu} T^{\mu i} &=& \partial_{0} T^{0 i} + \partial_{k} T^{k i} + \Gamma^{i}_{00} T^{00} + 2 
\Gamma^{i}_{0 j} T^{0 j} + \biggl(\Gamma^{0}_{0 0} + \Gamma^{k}_{0 k} \biggr) T^{i 0} 
\nonumber\\
&+& \Gamma^{i}_{j k} T^{j k} + \biggl(\Gamma^{0}_{j 0} + \Gamma^{k}_{j k} \biggr) T^{ij}, 
\label{T7}
\end{eqnarray}  
which can also be explicitly written, in ADM variables, as  
\begin{eqnarray}
\nabla_{\mu} T^{\mu 0} &=& \frac{1}{N^2}\biggl\{ \partial_{\tau} [ p\, s^2(y) + \rho\, c^2(y)]  + N^2 \partial_{k} \biggl[ \frac{p + \rho}{N} \, c(y) \,s(y) \,\hat{v}^{k}\biggr] 
\nonumber\\
&-& N K [ p\,s^2(y) + \rho\,c^2(y)]  + \biggl[3 \nabla_{k} N + N \Gamma_{k j}^{j}\biggr] (p + \rho) s(y) c(y) \,\hat{v}^{k}
\nonumber\\
&-& N K_{k j} \biggl[ (p + \rho) \hat{v}^{k} \, \hat{v}^{j} s^2(y) + p \gamma^{k j} + \Pi^{kj}\biggr]\biggr\},
\label{T9}\\
\nabla_{\mu} T^{\mu i} &=& \frac{1}{N} \partial_{\tau}[ ( p + \rho) s(y) c(y) \hat{v}^{i}]
+ \partial_{k}[ ( p + \rho) s^2(y) \hat{v}^{i} \hat{v}^{k} + p \gamma^{i k} + \Pi^{ki}] 
\nonumber\\
&+& \frac{\nabla^{i} N}{N} [ \rho\,c^2(y) + p\,s^2(y)]
\nonumber\\
&-&  2 K_{j}^{i} (p + \rho) s(y) \, c(y)\, \hat{v}^{j}  - K ( p + \rho) s(y) \, c(y)\, \hat{v}^{i} 
\nonumber\\
&+& \Gamma^{i}_{k j} [ ( p + \rho) \hat{v}^{k} \hat{v}^{j} s^2(y) + p\, \gamma^{k j} + \Pi^{k j}]
\nonumber\\
&+& \biggl( \frac{\nabla_{j} N}{N} + \Gamma^{k}_{j k}\biggr) [ (p + \rho) \hat{v}^{i} \hat{v}^{j} s^2(y) + p \gamma^{ij}  + \Pi^{i j}],
\label{T10}
\end{eqnarray}
where the shorthand notation $c(y) = \cosh{y}$ and $s(y) = \sinh{y}$ has been adopted and where  $\Pi^{ij}$ denotes the possible contribution of the anisotropic stress which has been included for completeness. Note that, in Eqs. (\ref{T9}) 
and (\ref{T10}) there is no potential ambiguity since, when the shift vector vanishes, $\Gamma_{ij}^{k}$ coincides 
with $^{(3)}\Gamma_{ij}^{k}$. This is not the case in general as the last equality of Eq. (\ref{ADM1c}) clearly shows.

Finally, the components 
of the electromagnetic energy-momentum tensor ${\mathcal T}_{\mu}^{\nu}$ become, in ADM variables,  
\begin{eqnarray}
{\mathcal T}_{0}^{0} &=& \frac{1}{8\pi \sqrt{\gamma}} \biggl[ q_{m n} \frac{ E^{m} E^{n} }{N} + \frac{\gamma_{n k} P_{i j m} \eta^{i j k}}{2} B^{m} B^{n} 
- \biggl( \frac{p_{n m}}{N} + \frac{\gamma_{nk} \eta^{i jk}}{2} Q_{i j m}\biggr) B^{n} E^{m} \biggr],
\label{T11}\\
{\mathcal T}_{i}^{0} &=& \frac{1}{4 \pi N \sqrt{\gamma}} \biggl[ 
Q_{i j k} E^{k} E^{j} - P_{i j k} B^{k} E^{j} \biggr],
\label{T12}\\
{\mathcal T}_{i}^{j} &=&  \frac{1}{4 \pi\sqrt{\gamma}}\biggl\{ 
\biggl( \frac{q_{i m}}{N} E^{m} E^{j} - \gamma_{a b} P_{i m n} \eta^{j m b} 
B^{n} B^{a} \biggr)  + \biggl( Q_{i m n} \gamma_{a b} \eta^{ j m b} E^{n} B^{a}
- p_{m i} \frac{B^{m} E^{j}}{N}\biggr) 
\nonumber\\
&+& \frac{1}{4} \delta_{i}^{j} \biggl[ P_{m n k} \gamma_{a b} \eta^{m n b} B^{k} B^{a}
+ \frac{2}{N} \biggl( p_{a b} B^{a} E^{b}-  q_{a b} E^{a} E^{b}\biggr)  
\nonumber\\
&-& Q_{m n k} \gamma_{a b} \eta^{m n  b} E^{k} B^{a}\biggr]\biggr\}.
\label{T13}
\end{eqnarray}
In Eqs. (\ref{T11}), (\ref{T12}) and (\ref{T13}) the following 
auxiliary tensors have been introduced, 
\begin{eqnarray}
q_{k i} &=& \frac{N_{k} N_{i} + \gamma_{i k} (N^2 - N_{m} N^{m})}{N \sqrt{\gamma}},
\qquad p_{k i} = \frac{N_{k} \gamma_{i j} \gamma_{m n} \eta^{m j n}}{\sqrt{\gamma}}, 
\label{T14}\\
Q_{i j k} &=& \frac{N_{j} \gamma_{i k} - N_{i} \gamma_{j k}}{N \sqrt{\gamma}},\qquad 
P_{i j k} = \frac{\gamma_{i m} \gamma_{j n} \gamma_{p k} \eta^{m n p}}{\sqrt{\gamma}}.
\label{T15}
\end{eqnarray}
In terms of the tensors given in Eqs. (\ref{T14}) and (\ref{T15})
the covariant components of the field strengths can be written as:
\begin{equation}
F_{0 i} = q_{k i} E^{k} - p_{k i} B^{k}, \qquad F_{i j}= Q_{i j k} E^{k} - P_{i j k} B^{k},
\label{T18}
\end{equation} 
while the controvariant components of the field strength have been already 
reported in Eqs. (\ref{EL2}) and (\ref{EL3}) in terms of ${\mathcal E}^{k} = (N/\sqrt{\gamma}) E^{k}$ and of ${\mathcal B}^{k} = (N/\sqrt{\gamma}) B^{k}$.
In the gauge where the shift vector vanishes, it is easy to show, within the 
present decomposition, that 
\begin{equation}
q_{k i} = \frac{N}{\sqrt{\gamma}} \gamma_{k i}, \qquad p_{k i} =0, 
\qquad Q_{i j k} =0,\qquad P_{i j k} = \frac{\gamma_{i m} \gamma_{j n} \gamma_{p k}}{\sqrt{\gamma}} \eta^{m n k}, 
\label{T19}
\end{equation}
and Eqs. (\ref{T11}), (\ref{T12}) and (\ref{T13}) 
reduce to:
\begin{eqnarray}
{\mathcal T}_{0}^{0} &=& \frac{\gamma_{m n}}{8 \pi \gamma}\biggl( E^{m} E^{n} 
+ B^{m} B^{n}\biggr), 
\label{T21}\\
{\mathcal T}_{i}^{0} &=&
- \frac{\gamma_{i m} \gamma_{j n} \gamma_{p k} 
\eta^{m n k}}{4\pi\,\gamma\, N} B^{p} E^{j},
\label{T22}\\
{\mathcal T}_{i}^{j} &=& \frac{1}{4 \pi \gamma} \biggl[ \gamma_{i m} 
E^{m} E^{j} + \gamma_{i m} B^{m} B^{j} - 
\frac{\delta_{i}^{j}}{2} \gamma_{m n} \biggl(E^{m} E^{n} + B^{m} B^{n}\biggr)\biggr].
\label{T23}
\end{eqnarray}
In terms of the shorthand notation employed in the bulk of the paper, the 
components reported in Eqs. (\ref{T21}), (\ref{T22}) and (\ref{T23}) 
can also be written as:
\begin{eqnarray}
{\mathcal T}_{0}^{0} &=& \frac{1}{8 \pi \gamma}\biggl( \vec{E}\cdot \vec{E} 
+ \vec{B}\cdot \vec{B}\biggr)  \equiv \rho_{\mathrm{B}} + \rho_{\mathrm{E}}, 
\label{T24}\\
{\mathcal T}_{i}^{0} &=&
- \frac{\gamma_{i m}}{4 \pi \gamma} (\vec{E} \times \vec{B})^{m},
\label{T25}\\
{\mathcal T}_{i}^{j} &=& \frac{1}{4 \pi \gamma} \biggl[ \gamma_{i m} 
E^{m} E^{j} + \gamma_{i m} B^{m} B^{j} - 
\frac{\delta_{i}^{j}}{2}  \biggl(\vec{E}\cdot \vec{E} 
+ \vec{B}\cdot \vec{B}\biggr)\biggr] 
\nonumber\\
&\equiv& - ( p_{\mathrm{E}} + p_{\mathrm{B}}) \delta_{i}^{j} + \Pi_{i}^{j}(E) + \Pi_{i}^{j}(B),
\label{T26}
\end{eqnarray}
where $p_{\mathrm{E}} = \rho_{\mathrm{E}}/3$ and $p_{\mathrm{B}} = \rho_{\mathrm{B}}/3$; $\Pi_{i}^{j}(E)$ and $\Pi_{i}^{j}(B)$ denote, respectively, the electric and 
the magnetic anisotropic stress.
\newpage
\renewcommand{\theequation}{B.\arabic{equation}}
\setcounter{equation}{0}
\section{Relations to Bardeen formalism}
\label{APPB}
The Bardeen formalism \cite{bardeen} is one the main tools customarily employed for a 
quantitative assessment of the impact of large-scale magnetic fields on the CMB anisotropies. 
It is therefore useful to describe the relation of the methods described in the present paper 
to the Bardeen approach. In the linearized theory of cosmological perturbations the fluctuations 
can be separated in scalar, vector and tensor modes as:
\begin{equation}
\delta g_{\mu\nu} = \delta_{\mathrm{s}} g_{\mu\nu} + \delta_{\mathrm{v}} g_{\mu\nu} +  \delta_{\mathrm{t}} g_{\mu\nu}.
\label{bard1}
\end{equation}
The scalar modes of the geometry are parametrized in terms of four independent functions. The vector modes 
are parametrized in terms of two pure vectors $W_{i}$ and $Q_{i}$ obeying $^{(3)}\nabla_{i} Q^{i} =0$ and 
$^{(3)}\nabla_{i} W^{i} =0$. Finally the tensor modes are parametrized in terms of a rank-two tensor $h_{ij}$ 
which is both divergenceless and traceless. Overall, before gauge fixing, the number of independent functions amounts to $10$. At nonlinear level, the decomposition of Eq. (\ref{bard1}) is meaningless but still the more general description discussed in this paper contains, as special cases, the gauge dependent approaches to the magnetized CMB anisotropies \cite{mg2,mg3}. For instance, the choice of the conformally Newtonian gauge (often 
dubbed longitudinal gauge)
corresponds to $N_{i} =0$, $N(\vec{x},\tau) = a(\tau) [ 1 + \phi(\vec{x},\tau)]$ and 
$\gamma_{ij}(\vec{x},\tau) = a^2(\tau)[1 - 2 \psi(\vec{x},\tau)] \delta_{ij}$. The extrinsic curvature 
and $\sqrt{\gamma}$ are therefore given by
\begin{equation}
K_{i}^{j} = \biggl[ - \frac{{\mathcal H}}{a} + \frac{1}{a} \biggl( \partial_{\tau} \psi + {\mathcal H} \phi \biggr)\biggr] \delta_{i}^{j},\qquad \sqrt{\gamma} = a^{3} ( 1 - 3 \psi).
\label{bard2}
\end{equation}
From Eqs. (\ref{ADM11}), (\ref{ADM12}) and (\ref{ADM13}) the various components of the einstein tensors 
can be obtained by keeping only the terms which are linear in the metric fluctuations.  For instance Eq. (\ref{ADM11}) implies 
\begin{eqnarray}
G_{0}^{0}(\vec{x},\tau) &=& \overline{G}_{0}^{0}(\tau) + \delta_{\mathrm{s}} G_{0}^{0}(\vec{x},\tau),
\label{bard3}\\
\overline{G}_{0}^{0}(\tau) &=& 3 \frac{{\mathcal H}^2}{a^2}, \qquad \delta_{\mathrm{s}} G_{0}^{0}(\vec{x},\tau) = 
\frac{2}{a^2} \biggl[ \nabla^2 \psi - 3 {\mathcal H} ({\mathcal H} \phi + \psi')\biggr].
\label{bard4}
\end{eqnarray}
It is useful to remark that, in Eq. (\ref{bard4}), $\biggl|\delta_{\mathrm{s}} G_{0}^{0}(\vec{x},\tau) \biggr| <  \overline{G}_{0}^{0}(\tau)$ as implied by the validity of the perturbative approximation. Conversely, in the gradient expansion, 
what matters is not the absolute magnitude of the perturbation in comparison with the background but rather 
the number of gradients defining the various orders of the expansion. Following the same procedure 
of Eq. (\ref{bard4}) the evolution equations in the longitudinal gauge can be explicitly obtained from Eqs. (\ref{ADM12}) 
and (\ref{ADM13}):
\begin{eqnarray}
&& \nabla^2 \psi - 3 {\mathcal H} ({\mathcal H} \phi + \partial_{\tau}\psi) = \frac{\ell_{\mathrm{P}}^2}{2} a^2 
[ \delta_{\mathrm{s}} \rho
+ \delta_{\mathrm{s}} \rho_{\mathrm{B}} + \delta_{\mathrm{s}} \rho_{\mathrm{E}}],
\label{bard5}\\
&& 2 \nabla^2 ({\mathcal H} \phi + \partial_{\tau} \psi)+ \ell_{\mathrm{P}}^2 \biggl[ (p_{\mathrm{t}} + \rho_{\mathrm{t}}) \theta_{\mathrm{t}} + \frac{\vec{\nabla}\cdot( \vec{E} \times \vec{B})}{4 \pi a^4}\biggr]=0,
\label{bard6}\\
&& 2 \biggl[ \partial_{\tau}^2 \psi + {\mathcal H}(\partial_{\tau} \phi + 2 \partial_{\tau} \psi) + ({\mathcal H}^2 + 2 \partial_{\tau} {\mathcal H})\phi + \frac{1}{2} \nabla^2(\phi - \psi) \biggr] \delta_{i}^{j}  +  \partial_{i}\partial^{j} (\phi - \psi) 
\nonumber\\
&& = \ell_{\mathrm{P}}^2 \biggl[ \biggl( \delta_{\mathrm{s}} p + \delta_{\mathrm{s}} p_{\mathrm{E}} + \delta_{\mathrm{s}} p_{\mathrm{B}} \biggr) \delta_{i}^{j} - \tilde{\Pi}_{i}^{j} - \Pi_{i}^{j}(E) - \Pi_{i}^{j}(B) \biggr],
\label{bard7}
\end{eqnarray}
where $\theta_{\mathrm{t}} = \vec{\nabla}\cdot \vec{v}_{\mathrm{t}}$. 
In analog terms the evolution equations can be obtained in a different gauge either by performing the appropriate gauge 
transformation on both sides of Eqs. (\ref{bard5}), (\ref{bard6}) and (\ref{bard7}) or by using again the general form 
of the Ricci (or Einstein) tensors reported in appendix \ref{APPA}. The synchronous gauge equations can be obtained 
by positing, 
\begin{equation}
N(\tau) = a(\tau),\qquad \gamma_{ij}(\vec{x},\tau) = a^2(\tau)[ ( 1 + 2 \xi) \delta_{ij} + 2 \partial_{i} \partial_{j} F].
\label{bard8}
\end{equation}
In the case of the synchronous gauge condition the extrinsic curvature and $\sqrt{\gamma}$ read
\begin{equation}
K_{i}^{j} = \frac{1}{a} \biggl[ - {\mathcal H} \delta_{i}^{j} - \partial_{\tau} \xi \delta_{i}^{j} - \partial_{i}\partial^{j} \partial_{\tau} F \biggr], \qquad \sqrt{\gamma} = a^3 ( 1 + 3 \xi + \nabla^2 F)
\label{bard9}
\end{equation}
Finally, not only the evolution 
of the scalar modes can be readily obtained but also those for the vector and for the tensor modes.  For instance, 
from Eq. (\ref{ADM9}),  in the gauge 
\begin{equation}
N(\tau) = a(\tau),\qquad \gamma_{ij} = a^2(\tau) 
\delta_{ij},\qquad N_{i}(\vec{x}, \tau) = a^2(\tau) Q_{i}(\vec{x},\tau),
\label{bard9a}
\end{equation}
the vector fluctuation of the Ricci tensor can be written:
\begin{eqnarray}
&& R_{i}^{j}(\vec{x},\tau) = \overline{R}_{i}^{j}(\tau) + \delta_{\mathrm{v}} R_{i}^{j}(\vec{x},\tau),
\nonumber\\
&& R_{i}^{j}(\tau) = - \frac{1}{a^2} \biggl(\partial_{\tau} {\mathcal H} + 2 {\mathcal H}^2\biggr)\delta_{i}^{j},
\nonumber\\
&& \delta_{\mathrm{v}} R_{i}^{j}(\vec{x},\tau) = \frac{1}{2 a^2} \biggl\{ \biggl[ \partial_{i} \partial_{\tau} Q^{j} + \partial^{j} \partial_{\tau} Q_{i} \biggr] + 2 {\mathcal H} \biggl[ \partial_{i} Q^{j} + \partial^{j}  Q_{i} \biggr]  \biggr\}.
\label{bard11}
\end{eqnarray}
In the bulk of the paper we shall not dwell on the derivation of the 
linearized results. It will be however understood that they are easily 
obtainable from the general expressions reported in appendix \ref{APPA} by going through the same steps 
outlined in the specific examples sketched in this appendix.  
\end{appendix}

\end{document}